# Nonlinear optics in 2D materials: from classical to quantum

Liuxin Gu[1], You Zhou[1,2*]

[1]Department of Materials Science and Engineering, University of Maryland, College Park, MD 20742, USA

[2]Maryland Quantum Materials Center, College Park, Maryland 20742, USA

*To whom correspondence should be addressed: youzhou@umd.edu

**Abstract.** Nonlinear optics has long been a cornerstone of modern photonics, enabling a wide array of technologies, from frequency conversion to the generation of ultrafast light pulses. Recent breakthroughs in two-dimensional (2D) materials have opened a frontier in this field, offering new opportunities for both classical and quantum nonlinear optics. These atomically thin materials exhibit strong light-matter interactions and large nonlinear responses, thanks to their tunable lattice symmetries, strong resonance effects, and highly engineerable band structures. In this paper, we explore the potential that 2D materials bring to nonlinear optics, covering topics from classical nonlinear optics to nonlinearities at the few-photon level. We delve into how these materials enable possibilities, such as symmetry control, phase matching, and integration into photonic circuits. The fusion of 2D materials with nonlinear optics provides insights into the fundamental behaviors of elementary excitations—such as electrons, excitons, and photons—in low-dimensional systems and has the potential to transform the landscape of next-generation photonic and quantum technologies.

## 1. Overview

The field of nonlinear optics examines the behaviors of light in media where the polarization density is not directly proportional to the electric field of light. Light entering such a nonlinear medium can alter the refractive index, generate new frequencies, or modify its own path through effects like self-focusing.[1,2] In addition, such nonlinearities can mediate interactions among photons, which become strong at high light intensities—typically requiring lasers to generate[3].

The study of nonlinear optics has significantly advanced our fundamental understanding of light-matter interactions under a strong optical field, enabling breakthroughs in nonlinear spectroscopy[4], frequency conversion[5], and the creation of non-equilibrium states in materials[6–8]. It has also propelled fields such as attosecond science[9], allowing scientists to probe the dynamics of atoms, molecules, and solids on ultrafast timescales shorter than one femtosecond. Meanwhile, nonlinear optics has also greatly expanded our capabilities to generate, manipulate, and detect light, forming the cornerstone of many modern optical technologies, such as ultrafast lasers[10,11], supercontinuum generation[12–14], and optical modulators[15–17], with broad applications in industries including telecommunications[18–21], spectroscopy[4,22,23], and bio-medical imaging[24,25].

Meanwhile, an emerging frontier in nonlinear optics explores the intersection with quantum optics, where both fields converge to enable groundbreaking quantum technologies[3] (**Fig. 1**). On the one hand, nonlinear optics facilitates the development of key quantum technologies, such as the generation of entangled photon pairs through spontaneous parametric down-conversion (SPDC)[26], which underpins quantum communication systems and networks[27,28]. On the other hand, remarkable progress in quantum optics has reduced the power requirements for nonlinear optical processes, reaching the single-photon level in meticulously engineered quantum systems[3,29]. Achieving such strong nonlinearity at the individual photon level allows for quantum control of light fields, opening avenues for unique applications such as single-photon switches, all-optical deterministic quantum logic, and highly efficient optical transistors[30–32].

Central to the science and technology of classical and quantum nonlinear optics is the challenge in developing materials with strong optical nonlinearity, higher efficiency, improved scalability, integration compatibility, and stability. Enhancing nonlinear coefficients in materials lowers the power needed for nonlinear effects, improving efficiency, accessibility, and scalability, but it proves difficult. Another major challenge is the integration nonlinear optical materials with linear optics platforms, such as photonic integrated circuits and metasurfaces[33,34]. This difficulty lies in achieving compatible growth and fabrication processes for nonlinear materials, such as lithium niobate ($LiNbO_3$), on substrates with distinct physical and chemical properties[5,35–37].

In this context, recent studies of two-dimensional (2D) materials, such as transition metal dichalcogenides (TMDs)[38], graphene[39], $CrI_3$[40], and black phosphorus[41], present promising opportunities for overcoming many challenges faced by traditional materials[42–46] (Fig.1). These atomically thin materials exhibit extraordinary properties, including readily controllable lattice symmetry, strong resonance effects, sizable interactions among quasiparticles, and tunable band gaps—all contributing to stronger light-matter interactions[47–49] and enhanced optical nonlinearity[50–54]. Additionally, 2D materials offer unprecedented flexibility in designing lattice and electronic structures by forming van der Waals heterostructures[42,55]. The heterostructures can then be readily integrated onto various photonic platforms, including photonic integrated circuits and metasurfaces[17,56–58]. As such, the study of nonlinear optical processes in 2D materials provides physical insights into electron, exciton, and phonon dynamics, opening new avenues for creating more efficient and compact optical devices for classical and quantum optical technologies.

This paper reviews the fundamental principles of nonlinear optics in 2D materials and explores their potential applications in both classical and quantum regimes. We first provide a brief overview of the mechanisms behind nonlinear optics, emphasizing the key properties of 2D materials that make them particularly intriguing. Next, we delve into various nonlinear optical processes in 2D materials, starting with second- and third-order nonlinearities and concluding with higher-order effects. We then focus on nonlinear optical phenomena at low photon levels, approaching the quantum regime, and discuss using nonlinear optics to engineer new states of matter through Floquet engineering. Finally, we offer an outlook on the field, discussing both the prospects and challenges in advancing the science and technology of nonlinear and quantum optics with 2D materials.

## 2. Introduction

### 2.1 Nonlinear optics mechanisms

The nonlinear optical effect occurs when the volume polarization $P(t)$ of a material does not linearly depend on the applied optical field $E(t)$. In particular, the optical response can be described by the following equation in the time domain:

$$P(t) = \epsilon_0\chi^{(1)}E(t) + \epsilon_0\chi^{(2)}E^2(t) + \epsilon_0\chi^{(3)}E^3(t) + \cdots\cdots\cdots \quad (1)$$

where $\epsilon_0$ is the permittivity in free space, $\chi^{(1)}$ is the materials linear susceptibility, and $\chi^{(n)}$refers to the *n*th order (n>1) nonlinear susceptibility. As seen from Equation (1), the strength of nonlinear optical processes substantially increases with stronger optical fields, and these higher-order processes enable the generation of different frequencies. More generally, equation (1) shall be expressed in the frequency domain and the $\chi^{(n)}(\omega)$ characterize the *n*th order (n>1) nonlinear susceptibility at frequency $\omega$.

Notably, $\chi^{(n)}(\omega)$ can be complex, allowing for the classification of nonlinear processes as either parametric or non-parametric. In parametric processes, only the real part of the susceptibility is relevant, meaning energy exchange occurs solely among interacting photons, with total photon energy conserved. The initial and final states of the material remain the same, without any photon absorption, or any exchange of momentum or angular momentum between the optical field. Since the material's quantum state remains unchanged, parametric processes can occur on extremely short timescales by the uncertainty principle ($\Delta t \approx \hbar/2\Delta E$). Examples of such include frequency conversion, such as second harmonic generation (SHG), via a virtual state.

A nonparametric process occurs when the material has a complex susceptibility with a nonzero imaginary component, leading to energy exchange between photons and the material itself. This often involves energy transfer between the ground and excited state, resulting in longer timescales than parametric process. Examples of nonparametric processes include saturable absorption (SA), two-photon absorption (TPA), and stimulated Raman scattering (SRS).

Phase matching is crucial for efficient frequency conversion in all nonlinear optical processes. While nonparametric processes inherently satisfy phase matching, this is not the case for parametric processes. For instance, in frequency doubling, two photons, each with a wavevector $\boldsymbol{k}_1$, combine to generate a photon with wavevector $\boldsymbol{k}_2$ at the harmonic frequency. Although the momentum is conserved, the chromatic dispersion of materials can induce a finite difference in their wavevectors as $\Delta\boldsymbol{k} = \boldsymbol{k}_2 - 2\boldsymbol{k}_1$. Ideally, efficient frequency conversion requires $\Delta k = 0$, ensuring all electric dipoles remain in phase and their emitted fields constructively add in the forward direction. Under these conditions, the harmonic field grows linearly with propagation length and its intensity quadratically. When $\Delta\boldsymbol{k}$ is nonzero, the harmonic field oscillates sinusoidally along the propagation length with a periodicity of $l_c \sim \frac{2\pi}{\Delta\boldsymbol{k}}$. The coherence length $l_c$, defines the maximum distance over which the harmonic field generated within the medium can constructively interfere with the field generated earlier. A finite coherence length inherently limits the efficiency of frequency conversion. Various techniques have been developed to overcome this limit, including birefringent, critical, and quasi-phase matching[59–61].

### 2.2 2D materials and opportunities for nonlinear optics

Over the past decade, two-dimensional (2D) materials have emerged as an exciting platform for photonics and optoelectronics[62]. These materials exhibit diverse structural, electronic, and optical properties, spanning metals, semimetals, semiconductors, insulators, and superconductors[63,64] (Fig. 2a). For instance, black phosphorus (BP) has a direct bandgap tunable by thickness[65,66], covering mid-infrared to ultraviolet spectra. The ability to stack individual layers into van der Waals heterostructures allows unprecedented flexibility in engineering electronic and optical responses[42,55], spurring significant advances in both linear and nonlinear optics.

2D materials possess unique attributes that make them promising for nonlinear optics[67–70]. Crystal symmetry, which plays a key role in even-order nonlinear processes, can be readily tailored in 2D heterostructures[71]. For example, transition metal dichalcogenides (TMDs) ($MX_2$, M = Mo, W; X = S, Se, Te) with strong optical response and bandgap in the visible and near-infrared regime[72–74], exhibit optical nonlinear behaviors dependent on their crystal, such as hexagonal 2H, rhombohedral 3R, or trigonal 1T' phases[75,76] (Fig. 2b). The stacking configurations and overall symmetry in fact can be artificially defined by controlling the twist angle during assembly of the structures to affect the nonlinear process[77]. Certain 2D materials with low structural symmetry (for example, BP) can lead to strong anisotropic optical properties[78], and in particular polarization-sensitive nonlinear optical processes[79,80].

Layers of 2D materials can be stacked with varying twist angles and material compositions to engineer different types of symmetry-breaking and nonlinear responses[77]. Controlling the twist angle proves important for realizing quasi-phase matching in such heterostructures. Additionally, twisting or lattice mismatching creates moiré superlattices, which can significantly influence optical nonlinearity. For example, in TMD moiré systems, the moiré potential can be strong enough to quantum confine excitons, leading to enhanced nonlinearity.

The reduced dimensionality of 2D materials can lead to relaxed phase matching, optical nonlinear response and strong light matter interactions[48,81,82]. In TMDs, for instance, weak screening and the relatively heavy carrier effective mass result in tightly bound excitons with large oscillator strengths[83,84]. This leads to a strong excitonic absorption and resonantly enhanced nonlinear effects[85,86](Fig. 2c). Additionally, the low-dimension nature promotes strong interactions between optical excitations, such as surface plasmon polaritons[87–89], phonon polaritons[89,90], and excitons polariton[91–93], mediating photon-photon interactions. Lastly, the atomic thickness of 2D materials — significantly thinner than the wavelength of light—results in negligible dispersive dephasing during propagation (Fig. 2d), inherently satisfying phase-matching conditions[75,94,95] and enhancing their suitability for nonlinear optical applications[96].

In addition, 2D materials can be readily controlled or engineered for various nonlinear optics applications. For example, the ability to control the twist angle between layers extends beyond simply modifying the overall symmetry. Careful design of the rotation between multiple layers in a heterostructure enables quasi-phase-matching. Moiré superlattices, generated at the twisting or lattice-mismatched 2D materials interface, can also strongly influence optical nonlinearity[97]. For instance, in TMD moiré systems, the moiré potential can be strong enough to quantum confine excitons[98,99], enhancing optical nonlinearity.

Other techniques, such as electrostatic gating[100,101] and strain tuning[102,103] , allow *in-situ* control of materials properly, enabling dynamic manipulation of nonlinear susceptibilities[104,105]. Furthermore, valley-dependent optical selection rules in 2D semiconductors enable coupling between circularly polarized light and specific valleys[106–108], with external magnetic fields or polarized optical pumping breaking valley degeneracy to achieve chiral nonlinearity[109,110].

Finally, 2D heterostructures can be integrated with virtually any substrate, regardless of lattice matching, making them highly suitable as nonlinear optical elements for photonic circuits and metasurfaces[56,111] , and an interface between optoelectronics and electronics (Fig. 2e). Techniques are being developed to create such structures with high quality and atomic precision, such at the back end-of-line (BEOL) processing stage.

Overall, the ability to control symmetry, exploit quantum confinement[112], harness strong resonance effects, and achieve ultrafast tunability[113], combined with the potential for integration, makes 2D materials an ideal candidate for next-generation nonlinear optics. Their applications include frequency conversion, optical modulation[15,114], compact high-performance photonic devices[115] , and quantum technology[44,45], highlighting their transformative potential in next-generation photonics.

## 3. Second-Order Nonlinearity

Second-order nonlinearity occurs when the induced polarization is proportional to the square of the electric field. Described by the second-order susceptibility tensor $\chi^{(2)}$, it gives rise to several important processes in nonlinear optics, including second harmonic generation (SHG), sum-frequency generation (SFG), and difference-frequency generation (DFG). These processes require materials with non-centrosymmetric geometry[116], since the inversion symmetry forces $\chi^{(2)}$ to vanish in centrosymmetric materials.

SHG, SFG, and DFG are frequency conversion processes enabled by the 2nd -order nonlinearity. In SHG process, a wave with frequency $\omega$ interacts with a medium possessing a finite $\chi^{(2)}$, generating a nonlinear polarization that produces a zero-frequency term and a second harmonic at $2\omega$. The conversion into the second harmonic can be highly efficient, reaching almost 100% conversion efficiency under optimal conditions. In contrast, the SFG and DFG occur when two waves at different frequencies $(\omega_1, \omega_2)$ interact to generate a third wave at a frequency $\omega_3$, where $\omega_3 = \omega_1 + \omega_2$ in SFG and $\omega_3 = \omega_1 - \omega_2$ in DFG, respectively. The DFG process is particularly useful for generating tunable light sources and entangled photon pairs, which will be discussed in more detail later. Crucially, in all these processes, efficient frequency conversion requires phase matching. Due to the dispersion of conventional nonlinear media materials, there is often a trade-off between the phase-matching bandwidth and conversion efficiency.

### 3.1 Second-Order Nonlinear effects in 2D materials

**3.1.1 Symmetry control.** Since second-order nonlinearity requires broken inversion symmetry, only specific types and stacking configurations of 2D materials can produce this nonlinear effect. Notably, monolayer transition metal dichalcogenides (TMDs) without any inversion symmetry have been explored extensively for these effects[76,117]. Intriguingly as the number of layers increases, different stacking configurations lead to different nonlinear behaviors. For example, in the 2H phase of TMDs, inversion symmetry is broken in materials with an odd number of layers but is restored when there is an even number of layers. This results in finite second-harmonic generation (SHG) in odd layers and zero SHG in even layers[118](Fig. 3a). In contrast, in the 3R phase (rhombohedral stacking), each monolayer is stacked in the same orientation, maintaining a non-centrosymmetric structure regardless of the number of layers. Consequently, in 3R stacking, the SHG intensity increases quadratically with the number of layers when the layer thickness is smaller than the coherence length, as the contribution from each layer adds up constructively[75,119,120]. The conversion efficiency reaches a maximum at the coherence length and oscillates as a function of propagation length with a period of $2l_c$[1].This observation points to the possibility of controlling the second-order nonlinear response by artificially stacking such layers rather than using naturally formed layers, which favors 2H stacking due to its thermodynamic stability.

The $\chi^{(2)}$ tensor reflects crystal symmetry and has material-specific frequency-dependence, making SHG a powerful tool for identifying the crystallographic orientation of materials with material selectivity in heterostructures. In these measurements, a linearly polarized laser is incident on the sample, and the SHG signal intensity is detected with the same linear polarization. By rotating the in-plane polarization, the SHG intensity can be mapped to determine the crystallographic orientation. For instance, in monolayer TMDs with $D_{3h}$ symmetry, the second-order nonlinear susceptibility follows $\chi^{(2)}_{xxx} = -\chi^{(2)}_{xyy} = -\chi^{(2)}_{yyx} = -\chi^{(2)}_{yxy}$, where x, y corresponds to the armchair and zigzag direction[118]. When the input laser polarization is parallel to the analyzer (the SHG signal polarization), the SHG intensity collected can be expressed as $I_{||} = I_0\, cos^2(3\phi)$. For the perpendicular case, the SHG intensity is: $I_{\perp} = I_0\, sin^2(3\phi)$, where $\phi$ is the angle between the input laser polarization and the armchair x direction[71]. As a result, the in-plane SHG intensity exhibits a six-fold symmetric pattern as the polarization is rotated in-plane (Figs. 3b-c).

**3.1.2 Resonant enhancement.** In 2D materials, strong resonance effect, such as excitons, can significantly influence their nonlinear response. In particular, the value of $\chi^{(2)}$ and SHG efficiency can be greatly enhanced when the SHG energy is close to any exciton resonance, due to the enhanced light-matter interaction[85,121]. In one study[85] (Fig. 3d), an enhanced SHG efficiency by a factor of 3 was observed in monolayer $WSe_2$ when scanning the two-photon laser energy across the excitonic spectrum. The strongest SHG intensity typically occurs at the 1s exciton state, when the 1s exciton is near the intermediate virtual transition levels. Since both the oscillator strength and energies of excitons can be tuned by doping[122], the resonantly enhanced SHG efficiency can also be modulated by gating. For instance, a four-fold reduction in the SHG intensity can be observed when electrostatically doping monolayer $WSe_2$[123].

**3.1.3 Electrical control of symmetry breaking.** An electric field can be used to break the inversion symmetry and induce a finite $\chi^{(2)}$ in materials without structural inversion symmetry breaking. For instance, in 2H-stacked bilayer $MoS_2$ with inversion symmetry, applying an electric field increases the SHG intensity to levels comparable to that in monolayer[124]. A stronger electric field leads to stronger electronic layer polarization, increasing the $\chi^{(2)}$ value.

This enhancement of $\chi^{(2)}$ in undoped materials is attributed to the interlayer exciton transitions, where the Coulomb-bound electrons and holes are spatially separated across two layers. Notably, interlayer excitons exhibit a static out-of-plane electric dipole, enabling their energies to be tuned by applied electric field via the Stark effect[100,125] (Fig. 3e). This also tunes the resonant energies at which $\chi^{(2)}$ enhancement is the most prominent. For example, the SHG intensity increases quadratically with the applied electric field, reaching a maximum enhancement by a factor of 25 at the interlayer exciton resonant energy of 0.17 MV/cm compared to the SHG intensity at zero electric field[124]. Similar effects have also been observed in 2H bilayer $WSe_2$[126] and 3R $MoS_2$[75].

In addition to forming dipolar interlayer excitons, an electric field can also break symmetry by polarizing doped carriers in structurally symmetric 2D materials, thereby inducing a finite SHG signal. For example, in a doped 2H-$WSe_2$ bilayer under an electric field, the layer-polarized free carriers can lead to a 40-fold enhancement in net SHG[127] (Fig. 3f-g).

Applying an electric-field gradient is another way to engineer symmetry. In bulk TMDs that are embedded inside a microcavity, the incident light can create an asymmetrically distributed electric field across the bulk TMD, breaking the inversion symmetry[128]. This asymmetry creates finite SHG by preventing the contributions from different layers from canceling out in the far field. The system's photonic density of state can be carefully engineered to enhance the SHG efficiency at the resonant polariton state.

**3.1.4 Nonlinear coefficient and conversion efficiency.**

Monolayer and few-layer TMDs can exhibit large nonlinear optical susceptibilities comparable to conventional nonlinear materials. For instance, in monolayers of $MoS_2$, $WS_2$, $WSe_2$, and $MoSe_2$, $\chi^{(2)}$ can reach the order of ~nm/V, even significantly higher than conventional nonlinear crystals, where $\chi^{(2)}$ is typically around pm/V[129]. However, the reported susceptibility in TMDCs varies widely, ranging from several pm/V to even hundreds of nm/V[68]. Several factors contribute to the variation in the reported values of susceptibilities, such as the materials quality, sample fabrication methods (e.g., exfoliated versus chemical vapor deposition (CVD) grown flakes)[130], local strain[131,132] and doping[133]. Furthermore, the nonlinear susceptibilities are highly wavelength-dependent and can be enhanced near optical resonances of materials, such as exciton energies, when nonparametric processes can become important. Therefore, it is crucial to consider these details when examining the nonlinear properties of TMDs (see Table 1 for summarized $\chi^{(2)}$ values in prominent 2D materials).

Despite the large nonlinear coefficient, the SHG efficiency in 2D materials is typically low due to their atomically thin nature. The intensity of SHG can be expressed as: $I_{2\omega} \propto$

$I_0^2 l^2 sinc^2(\frac{\Delta k \cdot l}{2})^1$, where $l$ is the propagation length. For instance, in $MoS_2$ [134], SHG efficiency at 810 nm is ~$10^{-7}$ with a Ti: Sapphire pump peak intensity of 10 GW $cm^{-2}$. In $WSe_2$ [123] at 1470 nm excitation with a pump peak intensity of 24 GW $cm^{-2}$, the estimated conversion efficiency for the on-resonance SHG (A exciton) is ~$10^{-10}$. This is in comparison with conventional NLO materials, such as thin film (600 nm) $LiNbO_3$, where the SHG conversion can reach a few percentages under similar pump peak intensity[135].

Further improving the conversion efficiency requires the increase of materials thickness so that signals within the coherence length can constructively interfere[136,137]. Stacking multiple layers of 2D materials provides a viable approach for enhancing SHG. However, chromatic dispersion must be carefully managed, and the TMD thickness must match the coherence length to ensure constructive interference (Fig. 4a)[108]. Another important factor is the measurement geometry. While conventional nonlinear materials are typically studied in transmission geometry, many experiments with 2D materials use reflection geometry. This simpler setup works well for nanoscale materials, where the interaction length is inherently limited by thickness. However, achieving higher efficiency in thicker samples requires characterization in transmission geometry, with careful phase-matching design and optimization of the photonic field, accounting for the multiple reflections at each interface.

One intriguing approach to enhance SHG is by twist engineering. For example, in multilayer hexagonal boron nitride (hBN), introducing a twist angle at the interface enhances SHG intensity. This arises from the broken mirror symmetry at the AB stacked interface[138] (Fig. 4b). In addition, such twist engineering can also lead to non-trivial polarization of SHG signals. For instance, it was predicted that in structures with a twist angle of 30°, a circular polarized SHG signal could be generated through linear polarized fundamental wave[139].

Engineering twist angles also provides a method to achieve quasi-phase matching. By alternating the crystalline orientation of stacked layers, the sign of $\chi^{(2)}$ is flipped periodically, enabling the SHG field to interfere constructively. This mechanism is conceptually analogous to the periodic poling used in conventional nonlinear crystal[140,141]. For instance, stacking two layers of 3R-$MoS_2$ with a thickness equal to the coherence length and a twist angle of 60° (or 180°/240° equivalently) results in a four-fold enhancement in SHG intensity compared to a single layer[142] (Fig. 4c).

Beyond periodic flipping at the coherence length, introducing multiple small twist angles across multiple layers before reaching the coherence length can effectively compensate for phase mismatch[138,143,144]. With such quasi-phase matching, a conversion frequency comparable to PPLN and BBO can be achieved in periodically poled 3R-MoS2 (Fig. 4d). Notably such 3R-$MoS_2$ system can be 10–100× thinner than traditional systems with similar performances[139,144].

### 3.1.5 SHG as a probe.

Polarization-resolved SHG, combined with excitation energy dependence, can offer critical insights into TMD heterostructures and moiré-related phenomena. For instance, in heterobilayers such as $MoS_2$/$WS_2$ and $MoS_2$/$WSe_2$, the interference of the light wave from the two layers can generate elliptical polarization of the overall SHG[139,145]. Analyzing the ellipticity of SHG signal provides critical information on the twist angle. Furthermore, a selectively exciting SHG that is resonant with a particular layer allows one to probe the crystallographic orientations of individual layers in a complex structure.

As SHG is highly sensitive to crystal symmetry, it can also detect edge states where translation symmetry is broken. The electronic structures at edges differ from those in bulk, contributing to enhanced second-order susceptibility and increased SHG intensity along edges[146,147]. In addition, edges with different atomic terminations, such as zigzag vs. armchair can have different resonant SHG due to their different electronic structure[147]. Therefore, edge SHG can serve as a sensitive probe for crystal structure, which can be particularly useful for *in-situ* characterization of crystal growth.

In 2D materials with valley-selective optical selection rules, such as TMDCs[148,149], polarization resolved SHG can have intriguing applications to valleytronics due to their ability to detect and control valley polarization[150]. For instance, SHG can be utilized as an efficient way to create excitons with a high degree of valley polarization. In particular, a circularly polarized fundamental wave generates second harmonic waves that are cross-polarized with almost unity valley polarization, due to the selection rule of the two-photon interband transitions[150–152]. (Figs. 4e-f).

Furthermore, resolving the polarization state of SHG signal allows for the measurement of valley polarization[151–153]. As an imbalance between the K and K' valley introduces additional $\chi^{(2)}$ term[154], modifying the SHG polarization state. By analyzing the ellipticity of the excitation and SHG waves, the valley polarization of carriers can be quantified[151–153](Fig. 4g). This SHG-enabled approach has been used to study temperature-dependent valley relaxation in TMDs[155] (Fig.4h). This capability provides critical information for valleytronics, such as the valley coherence properties and dynamics[156].

Lastly, SHG is a powerful probe for magnetic ordering in 2D materials. In centrosymmetric magnetic materials, non-centrosymmetric magnetic order can induce SHG[157]. This effect reflects the breaking of time-reversal symmetry in a time-noninvariant SHG process, as opposed to the typically dipole-allowed time-invariant SHG[157–159]. For example, the layered antiferromagnetic order in bilayer $CrI_3$ can be probed via SHG[158], which vanishes with an out-plane magnetic field and above the critical temperature at the antiferromagnetic-ferromagnetic phase to antiferromagnetic-paramagnetic phase transitions. This makes SHG a valuable tool for imaging magnetic domains and studying their dynamics[160,161].

### 3.1.6 OPA and OPO

The DFG process is widely used for generating coherent tunable light sources in processes such as optical parametric amplification (OPA) and optical parametric oscillation (OPO). During this process, the energy of the higher energy pump photon $h\omega_1$ transfer into the two lower energy states, creating an additional lower-energy photon, referred to as the signal, at $\omega_2$, and generating a third photon, called the idler, at $\omega_3$. With a strong pump, this process amplifies the intensity of the signal light in an OPA process. OPO, on the other hand, occurs when such a nonlinear system is placed inside an optical cavity, which is resonant with at least one of the signal and idler waves. This cavity feedback creates a continuous generation of signal and idler photons, turning the OPO into a coherent light source with high photon conversion efficiency.

OPA and OPO are impactful technologies for creating tunable coherent light sources across broad spectral ranges, as they amplify light within the transparency range of the nonlinear medium, provided phase-matching conditions are met. This can be widely used in quantum optics to generate squeezed coherent states of light and extend their operation bandwidth[162,163]. Currently, the bandwidth of OPO and OPA is largely constrained by the phase-matching requirements, such as the reliance on birefringence phase-matching[164,165].

With their atomic thickness and large susceptibilities, 2D materials provide unique opportunities to relax and engineer phase-matching conditions[165], enabling a wide range of applications. Early results demonstrate that in monolayer $MoSe_2$, the seed photon energy can be tuned over a broad range from 0.83 eV to 1.21 eV with a fixed pump energy (3.11 eV) while the idler photon energy changes from 2.28 eV to 1.90 eV[166] (Fig. 5a). This tuning range is broader than certain nonlinear materials, such as submicrometric periodically poled $KTiOPO_4$ (PPKTP) with a ~1.073 to 1.11 eV tuning range[165], though smaller than Ti: Sapphire waveguide amplifier with a tuning range of ~1.13 to 1.91 eV[167]. In another demonstration, inserting a monolayer of 1T'-$MoTe_2$ or 2H-$MoTe_2$ film into the cavity of a femtosecond OPO enabled effective modulation of the pulse spectral width and compression of pulse duration by a factor of 20[168].

Similar to SHG, the OPA and OPO efficiency can be also enhanced by increasing the number of layers in R-stacked multilayers, pointing to the new possibility of symmetry control[120]. These early results demonstrate the potential of 2D materials in OPA applications. By selecting materials with different bandgaps, it may be possible to achieve a larger transparency windows and wider tuning ranges, advancing the versatility of OPA and OPO systems[169].

### 3.1.7 Spontaneous Parametric Down Conversion (SPDC)

Spontaneous parametric down-conversion (SPDC) is a critical DFG process, where a pump photon spontaneously decays into two lower-energy photons: the signal and the idler. This process obeys both energy conservation as $\omega_1 = \omega_2 + \omega_3$ and momentum conservation $k_1 = k_2 + k_3$ but occurs without external control. The signal and idler photons, produced as a pair, can be entangled in various fashions, such as polarization, energy-time, and position-momentum entanglement. For instance, in a so-called type-II SPDC process, the signal and idler photons have orthogonal polarizations, forming a polarization-entangled Bell state[170–172]. As such, the SPDC process is essential for generating entangled photon pairs and single photons, making it a cornerstone of quantum optics and quantum information science[173].

Ultrathin 2D materials offer the potential to enable SPDC over a wide spectral range due to the relax phase-matching constraints. Recently, generation of correlated photons through SPDC was demonstrated in a 2D nonlinear material, niobium oxide dichloride ($NbOCl_2$)[46,174]. The photon statistics is characterized by the second-order correlation function $g^{(2)}(\tau)$ as a function of delay time $\tau$, which is defined as $g^{(2)}(\tau) = \frac{<N_s(t)N_i(t+\tau)>}{<N_s(t)><N_i(t+\tau)>}$, $N_{s,i}$ refers to the signal or idler photon numbers registered at the detectors. When pumped with a 3mW, 404nm continuous-wave laser, correlated photon pairs were generated, with a two-photon correlation peak $g^{(2)}(0)$ of ~25, indicating a temporal correlation between the generated photons[46]. Due to the low crystal symmetry in natural $NbOCl_2$, the efficiency of this process is highly polarization dependent, making it challenging to achieve well-defined polarization-entangled states. However, by leveraging van der waals engineering --- such as stacking two $NbOCl_2$ flakes with precise crystal alignment --- can overcome this limitation and enable the generation of entangled photon pairs.

In contrast, 2D materials with natural three-fold in-plane rotational symmetry, such as TMDs, enable the generation of well-defined polarization-entangled states[175]. For instance, when pumping with linear polarization direction along $x$- (zigzag) or $y$- (armchair), one can create two maximally entangled Bell states (Figs. 5b-c). These states correspond to $\phi^- = \frac{1}{\sqrt{2}}(|HH>_{s,i} - |VV>_{s,i})$ for $x$-polarized pump and $\psi^+ = \frac{1}{\sqrt{2}}(|HV>_{s,i} + |VH>_{s,i})$ for the $y$-polarized pump (Figs. 5d-e). By varying the pumping polarization angle, the generated state can be a superposition of the above two Bell states. Notably, thanks to the symmetry, the photon generation efficiency remains constant regardless of the pump polarization, a highly desirable feature for practical applications[175,176].

In addition, entangled photon pairs in the telecom regime (~1550nm) have been generated using multilayer 3R-$MoS_2$[175]. Experiments show a coincidence-to-accidental ratio (CAR) of 8.9$\pm$5.5 (where CAR = $g^{(2)}(0) - 1$), with low PL background noise under 5.6 mW of 788 nm excitation. By utilizing quasi-phase matching and the intrinsic cavity effect in the periodically poled 3R-MoS2 (with three poling periods and ~3.4 $\mu m$ thickness), the SPDC efficiency can be further enhanced, achieving a maximum CAR ratio of ~638 at the telecom wavelength[173,174](Figs. 5f-g). Though this surpasses the performance of most of the microscopic van der Waals SPDC source by two orders of magnitude[174,175], it's still low compared with that in conventional nonlinear crystals such as BBO or PPLN(Periodically Poled $LiNbO_3$ waveguide) and nanoscale metasurface[177–179]. Further increasing the propagation length in 2D materials could lead to additional improvements the SPDC efficiency[180,181].

Another promising avenue for improving nonlinear efficiency in 2D materials involves integrating them with metasurfaces. Because of their high refractive index, 2D materials themselves can be patterned into nonlinear metasurfaces that enhance light-matter interaction and nonlinear processes[182]. For instance, nanostructures designed to enable quasi-bound-state-in-the-continuum(q-BIC) resonance coupling to bulk 3R $MoS_2$ have demonstrated a three-order-of-magnitude enhancement in SHG compared to the original flake. These results highlight the potential of combining 2D materials with metasurface technologies to significantly boost nonlinear optical processes for more advanced quantum nanodevice[183,184].

## 4. Third-Order Nonlinearity

### 4.1 Parametric process

The third-order nonlinear process deals with the $\chi^{(3)}$nonlinearity term and is more common than second-order nonlinearity since it doesn't require broken inversion symmetry. Two important nonlinear optical processes can occur when the applied optical field is monochromatic: third-harmonic generation (THG) and the Kerr effect. In particular, with an input field of $E(t) = E\ cos\omega t$, the 3rd-order nonlinear dipole moment is given by[168]:

$$P^{(3)}(t) = \epsilon_0 \chi^{(3)} E^3(t) = \frac{1}{4}\epsilon_0 \chi^{(3)} E^3 \cos 3\omega t + \frac{3}{4}\epsilon_0 \chi^{(3)} E^3 \cos \omega t \quad (2)$$

The first term describes the THG process where three photons of frequency $\omega$ interact to generate one photon with a frequency of $3\omega$. The second term, proportional to $E(t)$, represents a change in the refractive index of the medium, experienced by the $\omega$ photons. This intensity-dependent refractive index change is also known as Kerr effect, and its effect on the refractive index can be described as:

$$n = n_0 + n_2 I \quad (3)$$

where $n_0$ is the linear refractive index, $I$ is the intensity of the incident beam ($\sim E^2$), and $n_2$ is the Kerr constant related to the real part of $\chi^{(3)}$:

$$n_2 = \frac{3}{4n_0^2\epsilon_0 c} Real(\chi^{(3)}) \quad (4)$$

The Kerr effect results in the modification of light propagation. For instance, when light with non-uniform intensity propagates in a material with a positive Kerr constant $n_2$, the regions of higher intensity experience a larger change in the refractive index. This causes the beam to bend toward the higher-intensity regions, making it self-focus.

### 4.1.1 Kerr and THG

The third-order nonlinear process exhibits ultrafast response in the femtosecond regime, and this has motivated the exploration of ultrafast all-optical switching and signal generation. The intensity-dependent refractive index change induced by the Kerr effect enables optical phase modulation, with 2D materials offering a significant advantage due to their large Kerr nonlinearity. For instance, monolayer TMDs exhibit Kerr constants on the order of $\sim 10^{-11}$ $m^2/W$[169], and graphene owns a large range of the Kerr constant from $\sim 10^{-11}$- $10^{-15}$ $m^2/W$ in telecommunication band[185]. These values are several orders of magnitude higher than that in bulk materials in silicon[186]($10^{-18}$ $m^2/W$) or silicon nitride[187]($10^{-19}$ $m^2/W$). In addition, resonant effects such as excitons can further enhance Kerr effects, which will be explored in more detail later in this review.

Third harmonic generation (THG) is an important nonlinear process with applications in short wavelength laser generation, spectroscopy[4], and imaging[187]. However, due to its lower efficiency compared to lower-order nonlinear processes, higher-order harmonic conversion requires greater pump power. It is reported that in monolayer $MoS_2$ the THG efficiency is $\sim 10^{-10}$ at the wavelength of 520nm[188]. This value scales with the number of layers in $MoS_2$(2H).

THG provides an additional approach to frequency generation. In a monolayer $MoS_2$, THG efficiency of $\sim 10^{-10}$ has been realized at the wavelength of 520nm[185]. This efficiency increases with the number of layers in 2H-stacked $MoS_2$. Similar to SHG processes, exciton resonances can significantly enhance THG efficiency. Such resonance enhancement has been observed in various 2D materials including $MoS_2$[189], $TaOI_2$ and $NbOI_2$ [190].

THG processes in 2D materials can be highly tunable and polarization dependent. In graphene, the linear dispersion and Dirac cone band structure result in strong third-order nonlinear susceptibility over a broadband spectral range. The THG efficiency in graphene can be modulated via chemical potential and electrostatic gating[191,192]. At finite doping, the absorption of photons is restricted to energies above $2E_F$ (where $E_F$ is the chemical potential) due to Pauli blocking[193], leading to a tunable third-order nonlinear response at specific wavelength.

In centrosymmetric materials with anisotropic structures, such as BP, strong polarization-dependent third order nonlinearity has been observed[194]. BP demonstrates several advantageous features for infrared optical modulators, including a large nonlinear absorption coefficient and low saturation density[195,196]. These properties make BP and similar anisotropic materials highly attractive for applications in nonlinear optics and optical modulation.

### 4.1.2 FWM

When the input field is polychromatic, the third-order nonlinearity enables a frequency conversion process called four waves mixing (FWM). FWM is highly useful for optical parametric amplification (OPA) and wavelength conversion applications. In non-degenerate four wave mixing (NDFWM) where $\omega_1 \neq \omega_2 \neq \omega_3$, three waves interact in the media to generate a fourth wave, at frequencies $\omega_4 = \omega_1 + \omega_2 + \omega_3$, $\omega_4 = \omega_1 + \omega_2 - \omega_3$, $\omega_4 = \omega_1 + \omega_3 - \omega_2$, or $\omega_4 = \omega_2 + \omega_3 - \omega_1$. In degenerate four-wave mixing (DFWM), there are three input fields at the same frequency $\omega_1 = \omega_2 = \omega_3$, which can propagate in different directions. With two strong beams and a counter-propagating weaker signal wave, a fourth wave is created at the same frequency ($\omega_4 = 2\omega - \omega$ ) but with a phase conjugated to the signal wave.

Four-wave mixing has been utilized as a spectroscopic technique to study exciton dynamics[197]. Specifically, in an optical two-dimensional Fourier-transform spectroscopy[198,199], three ultrafast pulses(in femtosecond) $\mathcal{E}_a^*, \mathcal{E}_b, \mathcal{E}_c$, are incident onto the sample to create a fourth FWM signal(Fig. 6a). The generated FWM and the excited three beams satisfy the phase-matching relationship as $\boldsymbol{k}_{FWM} = -\boldsymbol{k}_a + \boldsymbol{k}_b + \boldsymbol{k}_c$ .

By measuring the FWM intensity as a function of the time delay ($\tau_a$, $\tau_b$), the dephasing and relaxation dynamics of optical excitations can be quantified, which is typically challenging with linear spectroscopy. The process could be thought of as a photon echo. Specifically, the first pulse creates a coherent superposition of the ground state and excited state, which evolves freely within the first-time delay $\tau_a$, subject to the inhomogeneous broadening. The second and third reverse the state and phase accumulation during $\tau_c$, which cancels out the dephasing due to inhomogeneous broadening and produces a photon echo.

The 2D Fourier-transform spectra are obtained by transforming the measured signal $S(\tau_a, \tau_b, \tau_c)$ into the frequency domain $S(h\omega_A, \tau_b, h\omega_C)$. This analysis provides information about absorption, emission, and homogeneous versus inhomogeneous broadening[200](Fig.6b). This has been particularly useful for studying excitons in 2D materials, such as TMDs, which exhibit significant homogeneous broadening[201,202]. This enables the measurement of contributions from exciton-exciton[200,203], exciton-phonon[200,204] and charge-exciton[205] interactions to linewidth broadening. Furthermore, with circularly polarized pulses, valley decoherence can be measured by analyzing the FWM signal, providing critical insights into valley dynamics in 2D materials[206] (Figs. 6c-d).

## 4.2 Nonparametric process

Saturable absorption (SA) and two-photon absorption (TPA) are two examples of nonparametric third-order nonlinear effects. SA occurs when a high-intensity laser reduces the absorption coefficient and increases the transmission of the material. This effect is mainly caused by the finite number of available excited states in a material, such as the number of electronic states across the bandgap in a semiconductor. In the realm of single emitters such as single atoms or a single defect in a solid, a single photon can saturate the absorber, switching on and off the system's transmission. The relationship of the absorption coefficient $\alpha$ and the intensity $I$ can often be expressed as:

$$\alpha = \frac{\alpha_0}{1+\frac{I}{I_S}} \quad (5)$$

where $\alpha_0$ is the low-intensity absorption coefficient and $I_S$ is the saturation intensity. SA enables optical modulation of light intensities and is a key building block in photonics and optoelectronics applications, such as optical communication[206–208]. The ultrafast relaxation dynamics of carriers and excitons in 2D materials are particularly advantageous, enabling fast response and tuning, which has been used for passive mode-locking in a cavity for ultrashort laser pulse generation[207–209]. In addition, the operation wavelength window is highly dependent on the material's bandgap, covering a wide spectral range (Fig. 2a). For example, graphene-based SAs are notable for their wide wavelength operation range. However, further efforts are needed to engineer stronger absorption to achieve higher modulation depths.

Another third-order nonlinear absorption process is two-photon absorption (TPA). In TPA, two photons of the same frequency are absorbed simultaneously, promoting an electron from the ground state, such as the valence band, to an excited state, such as the conduction band. Because the total energy increase is equal to the combined photon energy, TPA occurs only when the energy of a single photon is greater than half of the bandgap of the nonlinear material. Unlike saturable absorption, the absorption coefficient in TPA grows with increasing input optical intensity.

Two-photon absorption allows for the detection of dark exciton states as it can excite optical species with different selection rules than linear spectroscopy. For instance, excitonic states with nonzero angular momentum, such as 2p and 3p excitons, usually inaccessible with linear optical detection, can be detected by two-photon excitation spectroscopy[210]. In monolayer $WS_2$, this enables the measurement of 2p and 3p exciton states[211]. Photoluminescence up-conversion has also been reported in monolayer $WSe_2$[212]. However, when real states such as excitons rather than virtual states are excited, other nonlinear processes such as Auger recombination[213], and multi phonon absorption[214] may start to play a role in addition to two-photon processes[215].

## 5. High Harmonic Generation

Beyond second and third-order nonlinearity, higher-order nonlinearity, such as high harmonic generation (HHG), can occur under an intense laser field approaching the ionization threshold of materials. HHG has been extensively studied in atomic and molecular gas over the past decades[216], and more recently in solids[217]. Unlike lower-order nonlinearity, HHG is a nonperturbative process, meaning that the harmonic yield does not scale as $|E|^q$ , where q is the order of the harmonics[217] as in perturbative nonlinear optics. Instead, a spectral plateau is often observed, over which the efficiency does not vary significantly with harmonic order. In addition, there is often a cut-off frequency, beyond which the harmonic intensity shows a steep drop[218]. The cutoff frequency can be roughly described by a semiclassical model where an electron, after escaping the ion under the influence of the laser field, recollides with the parent ion, resulting in the emission of high-energy photons[219]. HHG can be used generate ultrafast attosecond pulses, which opens up new avenues for studying various dynamics in solids at the attosecond timescale [9,220–222].

In 2D materials such as a monolayer $MoS_2$[223], HHG extending up to 13$^{th}$ order of the pump energy has been observed. The generated harmonic intensity varies as $\sim I^{3.3}$ for a range of harmonics, demonstrating its non-perturbative nature. In addition, in artificially stacked R-type multilayer $WS_2$, the HHG up to the 19$^{th}$ order is observed[224]. In this case, both the even- and odd-order harmonic intensity increase quadratically with the number of layers. As with other nonlinear processes, the phases of the generated waves play an important role in high-harmonic generation. Importantly, phase lock enables the formation of attosecond laser pulses[225]. In 2D materials, the study of phase matching and locking remain open areas for systematic exploration, offering opportunities to further optimize HHG efficiency in ultrathin systems.

## 6. Nonlinear optics in the few-photon regime

While traditional nonlinear optics relies on strong optical fields[226], the emerging field of quantum nonlinear optics seeks to achieve strong nonlinearity with lower power and ultimately even at the few-photon level. This has been a significant goal in both classical and nonclassical optics, as it enables energy-efficient devices such as optical switches, single-photon transistors, and gates[227–230]. However, photons, as the fundamental units of light, generally do not interact with each other unless under very specific conditions. One approach to facilitate strong photon-photon interactions is by integrating materials with strong nonlinearities. For example, systems such as single atoms[29,231], molecules[232], and atomic defects in solids[233], where the reflectance and absorption of the particle can be modified by a single photon with inherently strong nonlinearities. Such nonlinearity can be considered as, for instance, third-order nonlinearities such as saturable absorption with a saturation intensity corresponding to one photon per lifetime[234]. However, a major challenge to demonstrate single-photon nonlinearity lies in the deterministic coupling of light to these emitters, such as embedding them in optical resonators[29,230,231,235].

An alternative approach involves using extended optical excitations, such as excitons in semiconductors. Excitons have larger oscillator strengths, which lead to stronger coupling to light, making them a promising candidate for achieving the desired nonlinear interactions[201]. The exciton density-dependent resonance leads to third-order nonlinearity, which can be both dispersive and dissipative. The challenge, however, lies in the relatively weak nonlinearities due to the weak interactions among excitons that are charge-neutral. For instance, in 2D semiconductors such as monolayer $MoSe_2$ [202], the excitonic nonlinearity induced by the exciton-exciton interaction can be observed as exciton blueshift under pulsed laser excitation (Fig. 8a). The interaction strength, $g_{exc}$, is defined as the energy shift induced per unit area density of excitons. Rather generally, the exciton-exciton interaction $g_{exc}$, dominated by exchange interactions is on the order of $\sim E_B R_b^2$, where $E_B$ is the exciton binding energy and $R_b$ is the exciton Bohr radius. Notably, a smaller Bohr radius leads to tighter binding of excitons, which makes $g_{exc}$ more or less agnostic of materials, on the order of $\sim \mu eV \cdot \mu m^2$, whether in TMDs or GaAs quantum wells[236].

One way to enhance exciton nonlinearity is by spatially separating electrons and holes to form interlayer excitons (Fig. 8b). These interlayer excitons experience dipolar interactions, which can be stronger than those of intralayer excitons due to dipolar repulsion. In systems such as bilayer $WSe_2$[100], bilayer $MoS_2$[236], $WSe_2$/hBN/$MoSe_2$[237], and $MoSe_2$/hBN/$MoSe_2$[238], interlayer excitons exhibit a significant energy blueshift in photoluminescence (PL) or absorption as excitation power increases. The interaction strength increases with the electric dipole moment, following the parallel-plate capacitor model to the first order[239], making this nonlinear shift especially pronounced in interlayer excitons with a larger dipole moment, such as those separated by an additional hBN spacer.

It is also reported that the Coulomb interactions between excitons and free carriers can be much stronger than the exciton-exciton interactions and be used to facilitate large optical nonlinearity. For example (Fig. 8d,e), in homo-trilayer $WSe_2$[239], attractive Fermi polaron resonance also shows dramatic nonlinearity under both CW above band laser excitation and pulsed resonance excitations. The nonlinear interaction strength $g$ reaches as large as ~ $2\ meV \cdot \mu m^2$, significantly larger than exciton-exciton interactions. Interestingly, the nonlinearity only occurs when the trilayer is doped with holes, but not when it is intrinsic, or electron doped. This strong nonlinearity on the hole side was attributed to the valley polarization created by the exciton-carrier scattering, related to the near degenerate K and Γ valleys in such materials[240]. Such enhancement of interactions between optical excitations was also observed for the attractive polaron in $MoSe_2$ whose interaction strength is measured to be more than an order of magnitude stronger than those between bare excitons[241]. Crucially, this experiment characterizes the nonlinearity by exciting the material below its bandgap, which induces virtual populations and AC Stark shift (Fig. 8c). This helps to eliminate the complexity of measuring the population and interaction strength of excitons, such as dark excitons. In general, the gate-dependent response provides freedom for electrical tuning of the nonlinearity.

The nonlinear interaction strength to linewidth ratio can be further promoted when excitons are spatially confined. The quantum nonlinear optical regime is realized when the exciton blockade radius becomes larger than the confinement size. Therefore, quantum confining excitons such as using electrostatic gating[112,242], moiré superlattice[243,244], and remote potential imprinted from ferroelectric lattice become particularly interesting[245,246]. For instance, the trapped excitons in moiré superlattice can exhibit sharp linewidth[98], emit non-classical light[246], and experience strong nonlinearity[247].

Nonlinear optical effects have an important application in the Floquet engineering of quantum states. With ultrafast off resonant excitation, the exciton state becomes dressed by the incoming photon and forms a dressed state, leading to exciton energy blueshift (redshift) as in the optical Stark effect when the photon is red (blue) detuned[248,249] (Fig. 7a,b). In addition, the virtual excitons can interact with each other, leading to blueshift that depends on the exciton density[250]. The second effect is particularly strong when the pump field is at small detuning around the resonance and more virtual excitons are created. In general, such modification of excitons energy in a low virtual exciton density regime can be described as[241,249,250]:

$$\Delta E \approx \frac{A}{\Delta} + \frac{B}{\Delta^2} \tag{6}$$

where $\Delta$ is the energy detuning between photon and resonance. As the first term describes the exciton-photon interaction which is similar to the dressed-atom result, the second term describes the virtual exciton-exciton interaction. A is the term proportional to the intensity of applied photon field and the polarization matrix element between the ground state and A-exciton state. $\frac{B}{\Delta^2}$ is proportional to virtual exciton density and exciton-exciton interaction strength. Such control of the exciton energy by optical field forms a building block for Floquet engineering where new states of matter could emerge in a driven system.

Experimental modifications of the exciton energies have been reported in materials such as monolayer $WS_2$[249,251] and $MoSe_2$[252]. Valley-dependent optical selection rules enable valley-selective optical Stark effect[248] (Fig. 7c). A circularly polarized pump beam breaks the time-reversal symmetry and creates a valley imbalance of carriers at room temperature in monolayer $WS_2$[248,251].

In addition, the modification of exciton resonance can be quite substantial (Fig. 7d,e). With a strong optical field on the order of ~0.3V/nm(close to the ionization field of excitons), a blueshift at hundreds of meV can be observed in the 1S excitons of $MoSe_2$[251]. While these experiments often require a rather strong optical field, careful design of the optical environment, such as cavity engineering[253] and the use of plasmonic structures[254], can dramatically reduce the required power. Furthermore, some of these structures can be used to imprint spatially modulated optical field on 2D materials, opening exciting avenues for creating optical lattice with Floquet engineering and exploring novel correlated and topological physics[255].

### 6.1 Engineering nonlinearity through cavity enhancement

Given the intrinsically weak exciton-exciton interactions, various strategies are being pursued to enhance nonlinearity to the few-photon regime. A prominent approach involves integrating optical materials with cavities or waveguides[255]. For example, embedding TMDs into optical cavities leads to the formation of exciton-polaritons—a state that is half-light and half-matter—when the energy exchange between photons and excitons occurs faster than their respective decay rates (Fig. 9a). This strong light-matter coupling is typically characterized by the emergence of two polariton states with distinct energies, known as the lower and upper polaritons[256]. The coupling strength, also known as vacuum Rabi splitting, can be determined by observing the anti-crossing behavior between excitons and photons (Fig. 9b).

Crucially, polaritons often exhibit different nonlinear responses compared to bare excitons. In particular, the nonlinearity of polariton can be expressed as the susceptibility: $\chi^{(3)}(r) = \frac{16g_{exc-ph}^4}{|\Gamma|^2\Gamma}\frac{iU(r)}{\Gamma+iU(r)}$, where $g_{exc-ph}$ is the coupling strength between exciton and photon, $U(r)$ as the interaction strength between two excitons with a distance of r and $\Gamma$ as the linewidth. Meanwhile, the cavity modifies the linewidth of the excitons, often reducing the number of photons required to shift the resonance by a linewidth. Last but not least, with increasing pumping intensity, the vacuum Rabi splitting between the two polaritons also decreases, leading to the energy shift of polaritons[257,258]. The reduction in the splitting is due to the phase-space filling[258], where more exciton population leads to fewer excitons left to couple to the cavity photon, similar to saturable absorption. In fact, the Rabi splitting strength $\Omega$ depends on the polariton density: $\Omega = 2g\left(1 - \frac{n\pi R_b^2}{2}\right)$, leading to phase-space filling induced nonlinearity $g_{pol-pol} \propto a_b^2\hbar\Omega$[259]. At low density, the exchange interactions from the excitonic component dominate the polariton nonlinearity, resulting in a blue shift of both polariton branch resonances[260].

TMDs offer several appealing features for exploring the nonlinear interaction of polaritons in the strong light-matter coupling regime. With excitons in TMDs possessing large binding energies ranging from 200 to 500 meV and large oscillator strengths, strong coupling can persist even at room temperature with possibly enhanced exciton-mediated nonlinearity[261]. The nonlinearity of polaritons generated in TMDs varies significantly depending on the specific species coupled to the cavity photon. For instance (Fig. 9f), in monolayer $WSe_2$, the 2s exciton-polaritons exhibit a larger nonlinearity of approximately $46.4 \pm 13.9\ \mu eV \cdot \mu m^2$ , which is about four times larger than that of the 1s state[261]. This difference is expected, given the larger Bohr radius of 2s exciton. Biexciton polariton also exhibits nonlinear interaction boost[262]. Phase space filling was thought to play an important role in the measured nonlinear response in TMD polaritons. Similarly, trion-polariton in monolayer $MoSe_2$ microcavity exhibits enhancement on the nonlinearity strength of $\sim 37 \pm 3\ \mu eV \cdot \mu m^2$ at a low electron density and pump fluence[263] driven by the strong band-filling effect. However, the increased nonlinearity observed in 2s exciton- and trion-polaritons comes at the cost of reduced coupling strength between the optical excitation and the photons due to their weaker oscillator strength.

Another approach uses interlayer excitons with stronger dipolar repulsion, such as those in bilayer $MoS_2$, interlayer excitons[51](Fig. 9c). Such interlayer excitons in fact can have sufficient oscillator strength to enable strong coupling when the carrier tunneling is the materials is large enough to produce enough hybridization between interlayer and intralayer[51,263]. Combining this IX exciton with cavity photons forms the dipolar interlayer polariton, which has a ten-fold enhancement of this nonlinearity strength ($g_{IX}\sim\ 100\pm 2\ \mu eV\ \mu m^2$) compared with the conventional intralayer exciton (A) (~ 10 ) embedded in microcavity$g_A \pm 0.2 \mu eV\ \mu m^2$[51,264].

Similar to excitons in a cavity, it has been proposed[265] that one can significantly reduce the linewidth of the excitons by placing the monolayer $MoSe_2$ close to a partially reflecting mirror with a half-integer multiple of the exciton resonance wavelength. This could be understood as the destructive interference between the reflected light from the mirror and the emitted light from TMD substantially suppressing the radiative linewidth, leading to the formation of long-lived polariton, a hybrid of photon and exciton. Such reduction in linewidth reduces the number of photons required for shifting the exciton energy by a linewidth.

An intriguing set of recent experiments uses the moiré superlattice in 2D materials to demonstrate strong nonlinearity in exciton-polariton systems[266](Fig. 9d,e). With the moiré excitons confined within each site, they experience strong on-site interaction[267–269], which suggests that the exciton blockade can occur when average exciton-exciton density is close to the size of the moiré cell, on the order of a few to tens of nanometers, much larger than the blockade radius of bare excitons. Meanwhile, the exciton and photon can still couple collectively among all the cells, which maintains the system in a strong coupling regime.

With their recent development, it has become more feasible for this system to finally enter the quantum nonlinear optics regime when the polariton nonlinear interaction strength surpasses the linewidth. In such case, the optical response of the system is significantly altered by the presence of a single photon, leading to unprecedented phenomena and technologies in solids. One example is the so-called photon blockade[270,271], where a single polariton can inhibit the transmission of additional photons. Such photon blockade effect can be observed through photon antibunching and could open exciting avenues for realizing nonclassical light sources in quantum nonlinear optics[44,272–274].

## 7. Challenges and Outlook

This review has highlighted the immense potential of 2D materials in nonlinear optics, spanning both classical and quantum regimes. These materials offer unique opportunities to design novel heterostructures with well-controlled symmetry and band structures, leading to nonlinear optical responses often unattainable in bulk crystals. Their properties are highly tunable under external perturbations, unlocking functionalities for dynamically modulating and transducing nonlinear optical signals. Additionally, their atomically thin nature also introduces novel mechanisms for nonlinear processes, such as relaxing phase-matching conditions. The potential to integrate these heterostructures into various technological platforms, such as photonic integrated circuits, is particularly appealing for the development of nonlinear optical circuits[275–278].

Key challenges and opportunities remain to fully leverage the desired properties of 2D materials in developing nonlinear optical technologies. Addressing fundamental questions surrounding optical nonlinearity in 2D materials can lead to transformative insights. For instance, deeper understanding of interaction mechanisms, radiative vs. non-radiative decay, dephasing of optical excitations will unlock the potential of these materials[279]. Questions such as how to accurately determine the populations and interaction strengths of dark versus bright excitons remain open but are ripe for exploration through new theoretical models and advanced ultrafast or nanoscale measurement techniques. Additionally, the effects of lattice relaxation, disorder, and strain in heterostructures on exciton-phonon dynamics and optical nonlinearity remain not well understood and require further investigation[280]. Addressing these questions requires both the development of new theoretical models and novel instrumentation and methods capable of probing such phenomena at ultrafast timescales or ultrasmall length scales below the diffraction limit, which would, in turn, open a new frontier in the nonlinear optical science.

Beyond fundamental understanding, technical challenges in building efficient nonlinear optical devices provide a roadmap for innovation. Although the atomically thin nature of 2D materials limits interaction length and frequency conversion efficiency, solutions such as stacking monolayers into thicker structures, with careful engineering of the phase matching, offer promising directions[143,144]. Combining different materials such as active nonlinear materials and passive linear materials may compensate for phase mismatch[60]. Advances in scalable methods for direct growth or precise assembly of van der Waals heterostructures are crucial for overcoming current limitations and enabling the practical large-scale deployment of these devices.

Reassembling these monolayers into thicker layers could help address this limitation, but precise control over the material's symmetry and band structure is required to maintain a strong nonlinear response. Furthermore, as the thickness increases, phase-matching conditions become critical once again[281]. A potential solution involves fabricating multi-layer structures where active nonlinear materials are sandwiched between passive linear materials to compensate for phase mismatch. To create viable technologies[281], however, these complex structures must be scalable, potentially through direct growth methods.

For quantum nonlinear optics applications, enhancing the interaction strength among optical excitations and reducing their linewidth are critical goals. Spatial confinement of excitons can significantly lower the number of photons required to achieve blockade, which may be achieved via moiré superlattice[266,282–284], strain engineering[286], and electrostatic gating[287]. Additionally, engineering the resonance linewidth through cavity quantum electrodynamics, as well as improving material quality, can further reduce such photon threshold. The intrinsically short radiative lifetime of excitons[288,289], while challenging for photon statistics measurements, are advantageous for creating single-photon sources with high repetition rates and ultrafast optical switches. Developing innovative optical measurement techniques to capture these dynamics represents an exciting avenue for progress[290,291].

From the materials perspective, the application of 2D materials in optical devices requires the growth of wafer-scale high quality 2D materials. Overcoming challenges related to controlling crystallinity, defects, and stoichiometry in atomically thin films is critical[292]. The direct growth of phase-pure films on non-lattice-matched substrates, such as silicon and oxides, in a CMOS-compatible process is an area that demands significant effort[293,294]. In the realm of creating moiré or twisted heterostructures, current methods often result in large variations in interfacial structure, leading to device variability[293,294]. More scalable and reliable methods to fabricate van der Waals heterostructures with nanometric precision, either via transfer or growth, are needed for optical technologies[281,295].

The path forward requires multidisciplinary collaboration among physicists, materials scientists, optical engineers, and device designers. The integration of expertise across fields will accelerate progress toward scalable and efficient nonlinear optical technologies. With continued advancements in material growth, theoretical understanding, and device engineering, 2D materials have the potential to revolutionize nonlinear optics, opening new possibilities for applications in photonics, quantum technologies, and beyond.

In conclusion, while challenges remain, the outlook for 2D materials in nonlinear optics is overwhelmingly positive. Their unique properties, coupled with ongoing innovations, position them as transformative building blocks for the next generation of optical technologies.

**Acknowledgements**:

This research was supported by the U.S. Department of Energy DE-SC-0022885, DARPA HR0011-25-3-0303, and the National Science Foundation DMR-2145712.

**Reference:**

1. Boyd, R. W. *Nonlinear Optics*. (Elsevier, 2020).
2. Shen, Y. R. *Principles Of Nonlinear Optics*. (John Wiley and Sons (WIE), New York, NY, 1984).
3. Chang, D. E., Vuletić, V. & Lukin, M. D. Quantum nonlinear optics — photon by photon. *Nat. Photonics* **8**, 685–694 (2014).
4. Bloembergen, N. Nonlinear optics and spectroscopy. *Science* **216**, 1057–1064 (1982).
5. Keszler, D. A. Borates for optical frequency conversion. *Curr. Opin. Solid State Mater. Sci.* **1**, 204–211 (1996).
6. Eisert, J., Friesdorf, M. & Gogolin, C. Quantum many-body systems out of equilibrium. *Nat. Phys.* **11**, 124–130 (2015).
7. Hartmann, M. J. Quantum simulation with interacting photons. *J. Opt.* **18**, 104005 (2016).
8. Noh, C. & Angelakis, D. G. Quantum simulations and many-body physics with light. *Rep. Prog. Phys.* **80**, 016401 (2017).
9. Krausz, F. & Ivanov, M. Attosecond physics. *Rev. Mod. Phys.* **81**, 163–234 (2009).
10. Keller, U. Recent developments in compact ultrafast lasers. *Nature* **424**, 831–838 (2003).
11. Peccianti, M. *et al.* Demonstration of a stable ultrafast laser based on a nonlinear microcavity. *Nat. Commun.* **3**, 765 (2012).
12. Dudley, J. M., Genty, G. & Coen, S. Supercontinuum generation in photonic crystal fiber. *Rev. Mod. Phys.* **78**, 1135–1184 (2006).
13. Alfano, R. R. & Shapiro, S. L. Observation of self-phase modulation and small-scale filaments in crystals and glasses. *Phys. Rev. Lett.* **24**, 592–594 (1970).
14. Wright, L. G., Christodoulides, D. N. & Wise, F. W. Controllable spatiotemporal nonlinear effects in multimode fibres. *Nat. Photonics* **9**, 306–310 (2015).
15. Sun, Z., Martinez, A. & Wang, F. Optical modulators with 2D layered materials. *Nat. Photonics* **10**, 227–238 (2016).
16. Reed, G. T., Mashanovich, G., Gardes, F. Y. & Thomson, D. J. Silicon optical modulators. *Nat. Photonics* **4**, 518–526 (2010).
17. Klein, M. *et al.* 2D semiconductor nonlinear plasmonic modulators. *Nat. Commun.* **10**, 3264 (2019).
18. Agrawal, G. P. Nonlinear fiber optics: its history and recent progress [Invited]. *J. Opt. Soc. Am. B, JOSAB* **28**, A1–A10 (2011).
19. Feng, X. *et al.* Dispersion controlled highly nonlinear fibers for all-optical processing at telecoms wavelengths. *Opt. Fiber Technol.* **16**, 378–391 (2010).
20. Saitoh, K. & Koshiba, M. Highly nonlinear dispersion-flattened photonic crystal fibers for supercontinuum generation in a telecommunication window. *Opt. Express* **12**, 2027–2032 (2004).
21. Gu, T. *et al.* Regenerative oscillation and four-wave mixing in graphene optoelectronics. *Nat. Photonics* **6**, 554–559 (2012).
22. Chergui, M., Beye, M., Mukamel, S., Svetina, C. & Masciovecchio, C. Progress and prospects in nonlinear extreme-ultraviolet and X-ray optics and spectroscopy. *Nat Rev Phys* **5**, 578–596 (2023).

23. Dudovich, N., Oron, D. & Silberberg, Y. Single-pulse coherently controlled nonlinear Raman spectroscopy and microscopy. *Nature* **418**, 512–514 (2002).
24. Helmchen, F. & Denk, W. Deep tissue two-photon microscopy. *Nat. Methods* **2**, 932–940 (2005).
25. Parodi, V. *et al.* Nonlinear Optical Microscopy: From Fundamentals to Applications in Live Bioimaging. *Front. Bioeng. Biotechnol.* **8**, (2020).
26. Sultanov, V. *et al.* Tunable entangled photon-pair generation in a liquid crystal. *Nature* **631**, 294–299 (2024).
27. Dorfman, K. E., Schlawin, F. & Mukamel, S. Nonlinear optical signals and spectroscopy with quantum light. *Rev. Mod. Phys.* **88**, 045008 (2016).
28. Caspani, L. *et al.* Integrated sources of photon quantum states based on nonlinear optics. *Light Sci. Appl.* **6**, e17100 (2017).
29. Birnbaum, K. M. *et al.* Photon blockade in an optical cavity with one trapped atom. *Nature* **436**, 87–90 (2005).
30. Wang, J., Sciarrino, F., Laing, A. & Thompson, M. Integrated photonic quantum technologies. *Nat. Photonics* **14**, 273–284 (2019).
31. Miller, D. A. B. Are optical transistors the logical next step? *Nat. Photonics* **4**, 3–5 (2010).
32. Kimble, H. J. The Quantum Internet. *Nature* **453**, 1023–1030 (2008).
33. Wang, Y., Jöns, K. D. & Sun, Z. Integrated photon-pair sources with nonlinear optics. *Appl. Phys. Rev.* **8**, 011314 (2021).
34. Ferrera, M. *et al.* Low-power continuous-wave nonlinear optics in doped silica glass integrated waveguide structures. *Nature photonics* **2**, 737–740 (2008).
35. Zhang, M., Wang, C., Kharel, P., Zhu, D. & Lončar, M. Integrated lithium niobate electro-optic modulators: when performance meets scalability. *Optica* **8**, 652 (2021).
36. Wang, C. *et al.* Integrated lithium niobate electro-optic modulators operating at CMOS-compatible voltages. *Nature* **562**, 101–104 (2018).
37. Sasaki, T., Mori, Y., Yoshimura, M., Yap, Y. K. & Kamimura, T. Recent development of nonlinear optical borate crystals: key materials for generation of visible and UV light. *Mater. Sci. Eng. R Rep.* **30**, 1–54 (2000).
38. Wang, Q. H., Kalantar-Zadeh, K., Kis, A., Coleman, J. N. & Strano, M. S. Electronics and optoelectronics of two-dimensional transition metal dichalcogenides. *Nat. Nanotechnol.* **7**, 699–712 (2012).
39. Geim, A. K. Graphene: status and prospects. *Science* **324**, 1530–1534 (2009).
40. Gibertini, M., Koperski, M., Morpurgo, A. F. & Novoselov, K. S. Magnetic 2D materials and heterostructures. *Nat. Nanotechnol.* **14**, 408–419 (2019).
41. Xu, Y., Shi, Z., Shi, X., Zhang, K. & Zhang, H. Recent progress in black phosphorus and black-phosphorus-analogue materials: properties, synthesis and applications. *Nanoscale* **11**, 14491–14527 (2019).
42. Novoselov, K. S., Mishchenko, A., Carvalho, A. & Castro Neto, A. H. 2D materials and van der Waals heterostructures. *Science* **353**, aac9439 (2016).
43. Liu, X. & Hersam, M. C. 2D materials for quantum information science. *Nat. Rev. Mater.* **4**, 669–684 (2019).
44. Turunen, M. *et al.* Quantum photonics with layered 2D materials. *Nat. Rev. Phys.* **4**, 219–236 (2022).
45. Montblanch, A. R.-P., Barbone, M., Aharonovich, I., Atatüre, M. & Ferrari, A. C. Layered materials as a platform for quantum technologies. *Nat. Nanotechnol.* **18**, 555–571 (2023).

46. Guo, Q. *et al.* Ultrathin quantum light source with van der Waals NbOCl2 crystal. *Nature* **613**, 53–59 (2023).
47. Luo, Y. *et al.* Strong light-matter coupling in van der Waals materials. *Light Sci. Appl.* **13**, 203 (2024).
48. Wang, G. *et al.* Colloquium : Excitons in atomically thin transition metal dichalcogenides. *Rev. Mod. Phys.* **90**, 021001 (2018).
49. Wilson, N. P., Yao, W., Shan, J. & Xu, X. Excitons and emergent quantum phenomena in stacked 2D semiconductors. *Nature* **599**, 383–392 (2021).
50. Gu, L. *et al.* Giant optical nonlinearity of Fermi polarons in atomically thin semiconductors. *Nat. Photonics* **18**, 816–822 (2024).
51. Datta, B. *et al.* Highly nonlinear dipolar exciton-polaritons in bilayer MoS2. *Nat. Commun.* **13**, 6341 (2022).
52. Gu, J. *et al.* Enhanced nonlinear interaction of polaritons via excitonic Rydberg states in monolayer WSe2. *Nat. Commun.* **12**, 2269 (2021).
53. Wang, H.-X. *et al.* Quantum many-body simulation using monolayer exciton-polaritons in coupled-cavities. *J. Phys. Condens. Matter* **29**, 445703 (2017).
54. Englund, D. *et al.* Ultrafast photon-photon interaction in a strongly coupled quantum dot-cavity system. *Phys. Rev. Lett.* **108**, 093604 (2012).
55. Geim, A. K. & Grigorieva, I. V. Van der Waals heterostructures. *Nature* **499**, 419–425 (2013).
56. Li, P. *et al.* Infrared hyperbolic metasurface based on nanostructured van der Waals materials. *Science* **359**, 892–896 (2018).
57. Errando-Herranz, C. *et al.* Resonance fluorescence from waveguide-coupled, strain-localized, two-dimensional quantum emitters. *ACS Photonics* **8**, 1069–1076 (2021).
58. Zhang, Q. *et al.* Interface nano-optics with van der Waals polaritons. *Nature* **597**, 187–195 (2021).
59. Fejer, M. M., Magel, G. A., Jundt, D. H. & Byer, R. L. Quasi-phase-matched second harmonic generation: tuning and tolerances. *IEEE J. Quantum Electron.* **28**, 2631–2654 (1992).
60. Fiore, A., Berger, V., Rosencher, E., Bravetti, P. & Nagle, J. Phase matching using an isotropic nonlinear optical material. *Nature* **391**, 463–466 (1998).
61. Dunn, M. H. & Ebrahimzadeh, M. Parametric generation of tunable light from continuous-wave to femtosecond pulses. *Science* **286**, 1513–1518 (1999).
62. Mak, K. & Shan, J. Photonics and optoelectronics of 2D semiconductor transition metal dichalcogenides. *Nat. Photonics* **10**, 216–226 (2016).
63. Akinwande, D. *et al.* Graphene and two-dimensional materials for silicon technology. *Nature* **573**, 507–518 (2019).
64. Young, A. F. & Kim, P. Electronic transport in graphene heterostructures. *Annu. Rev. Condens. Matter Phys.* **2**, 101–120 (2011).
65. Higashitarumizu, N. *et al.* Anomalous thickness dependence of photoluminescence quantum yield in black phosphorous. *Nat. Nanotechnol.* **18**, 507–513 (2023).
66. Ling, X., Wang, H., Huang, S., Xia, F. & Dresselhaus, M. S. The renaissance of black phosphorus. *Proc. Natl. Acad. Sci. U. S. A.* **112**, 4523–4530 (2015).
67. Dogadov, O., Trovatello, C., Yao, B., Soavi, G. & Cerullo, G. Parametric nonlinear optics with layered materials and related heterostructures. *Laser Photon. Rev.* **16**, 2100726 (2022).
68. Autere, A. *et al.* Nonlinear optics with 2D layered materials. *Adv. Mater.* **30**, e1705963 (2018).
69. Xie, Z., Zhao, T., Yu, X. & Wang, J. Nonlinear Optical Properties of 2D Materials and their Applications. *Small* **20**, e2311621 (2024).

70. Shi, J., Feng, S., He, P., Fu, Y. & Zhang, X. Nonlinear optical properties from engineered 2D materials. *Molecules* **28**, 6737 (2023).
71. Wang, Y., Xiao, J., Yang, S., Wang, Y. & Zhang, X. Second harmonic generation spectroscopy on two-dimensional materials [Invited]. *Opt. Mater. Express* **9**, 1136 (2019).
72. Manzeli, S., Ovchinnikov, D., Pasquier, D., Yazyev, O. V. & Kis, A. 2D transition metal dichalcogenides. *Nat. Rev. Mater.* **2**, 17033 (2017).
73. Splendiani, A. *et al.* Emerging photoluminescence in monolayer MoS2. *Nano Lett.* **10**, 1271–1275 (2010).
74. Mak, K. F., Lee, C., Hone, J., Shan, J. & Heinz, T. F. Atomically thin $MoS_2$: a new direct-gap semiconductor. *Phys. Rev. Lett.* **105**, 136805 (2010).
75. Zhao, M. *et al.* Atomically phase-matched second-harmonic generation in a 2D crystal. *Light Sci. Appl.* **5**, e16131 (2016).
76. Shi, J. *et al.* 3R MoS2 with broken inversion symmetry: A promising ultrathin nonlinear optical device. *Adv. Mater.* **29**, (2017).
77. Withers, F. *et al.* Light-emitting diodes by band-structure engineering in van der Waals heterostructures. *Nat. Mater.* **14**, 301–306 (2015).
78. Xia, F., Wang, H. & Jia, Y. Rediscovering black phosphorus as an anisotropic layered material for optoelectronics and electronics. *Nat. Commun.* **5**, 4458 (2014).
79. Yuan, H. *et al.* Polarization-sensitive broadband photodetector using a black phosphorus vertical p-n junction. *Nat. Nanotechnol.* **10**, 707–713 (2015).
80. Li, D. *et al.* Polarization and thickness dependent absorption properties of black phosphorus: New saturable absorber for ultrafast pulse generation. *Sci. Rep.* **5**, 15899 (2015).
81. Ma, E. Y. *et al.* The Reststrahlen effect in the optically thin limit: A framework for resonant response in thin media. *Nano Lett.* **22**, 8389–8393 (2022).
82. Munkhbat, B., Wróbel, P., Antosiewicz, T. J. & Shegai, T. O. Optical constants of several multilayer transition metal dichalcogenides measured by spectroscopic ellipsometry in the 300-1700 nm range: High index, anisotropy, and hyperbolicity. *ACS Photonics* **9**, 2398–2407 (2022).
83. Chernikov, A. *et al.* Exciton binding energy and nonhydrogenic Rydberg series in monolayer WS(2). *Phys. Rev. Lett.* **113**, 076802 (2014).
84. Li, Y. *et al.* Measurement of the optical dielectric function of monolayer transition-metal dichalcogenides:MoS2,MoSe2,WS2, andWSe2. *Phys. rev. B* **90**, 205422 (2014).
85. Wang, G. *et al.* Giant enhancement of the optical second-harmonic emission of WSe(2) monolayers by laser excitation at exciton resonances. *Phys. Rev. Lett.* **114**, 097403 (2015).
86. Soh, D. B. S., Rogers, C., Gray, D. J., Chatterjee, E. & Mabuchi, H. Optical nonlinearities of excitons in monolayer MoS2. *Phys. Rev. B.* **97**, 1–21 (2018).
87. Ni, G. X. *et al.* Plasmons in graphene moiré superlattices. *Nat. Mater.* **14**, 1217–1222 (2015).
88. Fei, Z. *et al.* Gate-tuning of graphene plasmons revealed by infrared nano-imaging. *Nature* **487**, 82–85 (2012).
89. Caldwell, J. D. *et al.* Low-loss, infrared and terahertz nanophotonics using surface phonon polaritons. *Nanophotonics* **4**, 44–68 (2015).
90. Dai, S. *et al.* Tunable phonon polaritons in atomically thin van der Waals crystals of boron nitride. *Science* **343**, 1125–1129 (2014).
91. Dufferwiel, S. *et al.* Exciton-polaritons in van der Waals heterostructures embedded in tunable microcavities. *Nat. Commun.* **6**, 8579 (2015).

92. Gan, X. *et al.* Controlling the spontaneous emission rate of monolayer MoS2 in a photonic crystal nanocavity. *Appl. Phys. Lett.* **103**, 181119 (2013).
93. Liu, X. *et al.* Strong light–matter coupling in two-dimensional atomic crystals. *Nat. Photonics* **9**, 30–34 (2015).
94. Xu, X. *et al.* Towards compact phase-matched and waveguided nonlinear optics in atomically layered semiconductors. *Nat. Photonics* **16**, 698–706 (2022).
95. Ciattoni, A., Marini, A., Rizza, C. & Conti, C. Phase-matching-free parametric oscillators based on two-dimensional semiconductors. *Light Sci. Appl.* **7**, 5 (2018).
96. You, J. W., Bongu, S. R., Bao, Q. & Panoiu, N. C. Nonlinear optical properties and applications of 2D materials: theoretical and experimental aspects. *Nanophotonics* **8**, 63–97 (2018).
97. Du, L. *et al.* Nonlinear physics of moiré superlattices. *Nat. Mater.* **23**, 1179–1192 (2024).
98. Seyler, K. L. *et al.* Signatures of moiré-trapped valley excitons in MoSe2/WSe2 heterobilayers. *Nature* **567**, 66–70 (2019).
99. Baek, H. *et al.* Highly energy-tunable quantum light from moiré-trapped excitons. *Sci. Adv.* **6**, eaba8526 (2020).
100. Wang, Z., Chiu, Y.-H., Honz, K., Mak, K. F. & Shan, J. Electrical tuning of interlayer exciton gases in WSe2 bilayers. *Nano Lett.* **18**, 137–143 (2018).
101. Hagel, J., Brem, S. & Malic, E. Electrical tuning of moiré excitons in $MoSe_2$ bilayers. *2d Mater.* **10**, 014013 (2023).
102. Aslan, B., Deng, M. & Heinz, T. F. Strain tuning of excitons in monolayer WSe2. *Phys. Rev. B.* **98**, 115308 (2018).
103. Bai, Y. *et al.* Excitons in strain-induced one-dimensional moiré potentials at transition metal dichalcogenide heterojunctions. *Nat. Mater.* **19**, 1068–1073 (2020).
104. Guan, Z. *et al.* Giant second-order susceptibility in monolayer WSe2 via strain engineering. *arXiv* (2024).
105. Glazov, M. M., Golub, L. E., Wang, G., Marie, X. & Amand, T. Intrinsic exciton-state mixing and nonlinear optical properties in transition metal dichalcogenide monolayers. *Phys. Rev. B.* **95**, 035311 (2017).
106. Mak, K. F., He, K., Shan, J. & Heinz, T. F. Control of valley polarization in monolayer MoS2 by optical helicity. *Nat. Nanotechnol.* **7**, 494–498 (2012).
107. Zeng, H., Dai, J., Yao, W., Xiao, D. & Cui, X. Valley polarization in MoS2 monolayers by optical pumping. *Nat. Nanotechnol.* **7**, 490–493 (2012).
108. Sallen, G. *et al.* Robust optical emission polarization in MoS2monolayers through selective valley excitation. *Phys. Rev. B.* **86**, 081301 (2012).
109. Alireza Taghizadeh, T. G. P. Nonlinear optical selection rules of excitons in monolayer transition metal dichalcogenides. *Phys. Rev. B* **99**, 1–11 (2019).
110. Zhang, Y. *et al.* Coherent modulation of chiral nonlinear optics with crystal symmetry. *Light Sci. Appl.* **11**, 216 (2022).
111. Muhammad, N., Chen, Y., Qiu, C.-W. & Wang, G. P. Optical bound states in continuum in MoS2-based metasurface for directional light emission. *Nano Lett.* **21**, 967–972 (2021).
112. Thureja, D. *et al.* Electrically tunable quantum confinement of neutral excitons. *Nature* **606**, 298–304 (2022).
113. Jin, C. *et al.* Ultrafast dynamics in van der Waals heterostructures. *Nat. Nanotechnol.* **13**, 994–1003 (2018).

114. Yu, S., Wu, X., Wang, Y., Guo, X. & Tong, L. 2D materials for optical modulation: Challenges and opportunities. *Adv. Mater.* **29**, 1606128 (2017).
115. Youngblood, N. & Li, M. Integration of 2D materials on a silicon photonics platform for optoelectronics applications. *Nanophotonics* **6**, 1205–1218 (2016).
116. Shen, Y. R. Surface properties probed by second-harmonic and sum-frequency generation. *Nature* **337**, 519–525 (1989).
117. Fan, X. *et al.* Broken Symmetry Induced Strong Nonlinear Optical Effects in Spiral WS 2 Nanosheets. *ACS Nano* **11**, 4892–4898 (2017).
118. Malard, L. M., Alencar, T. V., Barboza, A. P. M., Mak, K. F. & de Paula, A. M. Observation of intense second harmonic generation from MoS2atomic crystals. *Phys. Rev. B.* **87**, 201401 (2013).
119. Liu, F. *et al.* Disassembling 2D van der Waals crystals into macroscopic monolayers and reassembling into artificial lattices. *Science* **367**, 903–906 (2020).
120. Qin, B. *et al.* Interfacial epitaxy of multilayer rhombohedral transition-metal dichalcogenide single crystals. *Science* **385**, 99–104 (2024).
121. Lafrentz, M. *et al.* Second-harmonic generation spectroscopy of excitons in ZnO. *Phys. Rev. B* **88**, 235207 (2013).
122. Wang, K. *et al.* Electrical control of charged carriers and excitons in atomically thin materials. *Nat. Nanotechnol.* **13**, 128–132 (2018).
123. Seyler, K. L. *et al.* Electrical control of second-harmonic generation in a WSe2 monolayer transistor. *Nat. Nanotechnol.* **10**, 407–411 (2015).
124. Shree, S. *et al.* Interlayer exciton mediated second harmonic generation in bilayer MoS2. *Nat. Commun.* **12**, 6894 (2021).
125. Leisgang, N. *et al.* Giant Stark splitting of an exciton in bilayer MoS2. *Nat. Nanotechnol.* **15**, 901–907 (2020).
126. Klein, J. *et al.* Electric-field switchable second-harmonic generation in bilayer MoS2 by inversion symmetry breaking. *Nano Lett.* **17**, 392–398 (2017).
127. Cha, S. *et al.* Enhancing resonant second-harmonic generation in bilayer WSe2 by layer-dependent exciton-polaron effect. *Nano Lett.* **24**, 14847–14853 (2024).
128. Khatoniar, M. *et al.* Relaxing symmetry rules for nonlinear optical interactions in van der Waals materials via strong light–matter coupling. *ACS Photonics* **9**, 503–510 (2022).
129. Nikogosyan, D. N. *Nonlinear Optical Crystals: A Complete Survey*. (Springer, New York, NY, 2005).
130. Zhou, R., Krasnok, A., Hussain, N., Yang, S. & Ullah, K. Controlling the harmonic generation in transition metal dichalcogenides and their heterostructures. *Nanophotonics* **11**, 3007–3034 (2022).
131. Beach, K., Lucking, M. C. & Terrones, H. Strain dependence of second harmonic generation in transition metal dichalcogenide monolayers and the fine structure of the C exciton. *Phys. Rev. B.* **101**, (2020).
132. Rhim, S. H., Kim, Y. S. & Freeman, A. J. Strain-induced giant second-harmonic generation in monolayered 2*H*-MoX2 (X = S, Se, Te). *Appl. Phys. Lett.* **107**, 241908 (2015).
133. Tan, S. J. R. *et al.* Chemical Stabilization of 1T′ Phase Transition Metal Dichalcogenides with Giant Optical Kerr Nonlinearity. *J. Am. Chem. Soc.* **139**, 2504–2511 (2017).
134. Kumar, N. *et al.* Second harmonic microscopy of monolayer MoS 2. *Phys. Rev. B* **87**, 161403 (2013).

135. Zhao, Y., Chen, Z., Wang, C., Yang, Y. & Sun, H. B. Efficient second-and higher-order harmonic generation from LiNbO 3 metasurfaces. *Nanoscale* **15**, 12926–12932 (2023).
136. Martorell, J., Vilaseca, R. & Corbalán, R. Second harmonic generation in a photonic crystal. *Appl. Phys. Lett.* **70**, 702–704 (1997).
137. Bloembergen, N., Chang, R. K., Jha, S. S. & Lee, C. H. Optical second-harmonic generation in reflection from media with inversion symmetry. *Phys. Rev.* **174**, 813–822 (1968).
138. Yao, K. *et al.* Enhanced tunable second harmonic generation from twistable interfaces and vertical superlattices in boron nitride homostructures. *Sci. Adv.* **7**, eabe8691 (2021).
139. Paradisanos, I. *et al.* Second harmonic generation control in twisted bilayers of transition metal dichalcogenides. *Phys. Rev. B.* **105**, (2022).
140. Wang, C. *et al.* Ultrahigh-efficiency wavelength conversion in nanophotonic periodically poled lithium niobate waveguides. *Optica* **5**, 1438–1441 (2018).
141. Niu, Y. *et al.* Research progress on periodically poled lithium niobate for nonlinear frequency conversion. *Infrared Phys. Technol.* **125**, 104243 (2022).
142. Tang, Y. *et al.* Quasi-phase-matching enabled by van der Waals stacking. *Nat. Commun.* **15**, 9979 (2024).
143. Trovatello, C. *et al.* Quasi-phase-matched up- and down-conversion in periodically poled layered semiconductors. *Nat. Photonics* 1–9 (2025).
144. Hong, H. *et al.* Twist phase matching in two-dimensional materials. *Phys. Rev. Lett.* **131**, 233801 (2023).
145. Kim, W., Ahn, J. Y., Oh, J., Shim, J. H. & Ryu, S. Second-harmonic Young's interference in atom-thin heterocrystals. *Nano Lett.* **20**, 8825–8831 (2020).
146. Yin, X. *et al.* Edge nonlinear optics on a $MoS_2$ atomic monolayer. *Science* **344**, 488–490 (2014).
147. Lin, K.-I. *et al.* Atom-dependent edge-enhanced second-harmonic generation on MoS2 monolayers. *Nano Lett.* **18**, 793–797 (2018).
148. Schaibley, J. R. *et al.* Valleytronics in 2D materials. *Nat. Rev. Mater.* **1**, 1–15 (2016).
149. Mak, K. F., Xiao, D. & Shan, J. Light–valley interactions in 2D semiconductors. *Nat. Photonics* **12**, 451–460 (2018).
150. Klimmer, S. *et al.* All-optical polarization and amplitude modulation of second-harmonic generation in atomically thin semiconductors. *Nat. Photonics* **15**, 837–842 (2021).
151. Ho, Y. W. *et al.* Measuring Valley Polarization in Two-Dimensional Materials with Second-Harmonic Spectroscopy. *ACS Photonics* **7**, 925–931 (2020).
152. Herrmann, P. *et al.* Nonlinear valley selection rules and all-optical probe of broken time-reversal symmetry in monolayer WSe2. *Nat. Photonics* 1–7 (2025).
153. Herrmann, P. *et al.* Nonlinear all-optical coherent generation and read-out of valleys in atomically thin semiconductors. *Small* **19**, e2301126 (2023).
154. Hipolito, F. & Pereira, V. M. Second harmonic spectroscopy to optically detect valley polarization in 2D materials. *2d Mater.* **4**, 021027 (2017).
155. Mouchliadis, L. *et al.* Probing valley population imbalance in transition metal dichalcogenides via temperature-dependent second harmonic generation imaging. *Npj 2D Mater. Appl.* **5**, 1–9 (2021).
156. Lundt, N. *et al.* Optical valley Hall effect for highly valley-coherent exciton-polaritons in an atomically thin semiconductor. *Nat. Nanotechnol.* **14**, 770–775 (2019).
157. Fiebig, M., Pavlov, V. V. & Pisarev, R. V. Second-harmonic generation as a tool for studying electronic and magnetic structures of crystals: review. *J. Opt. Soc. Am. B* **22**, 96 (2005).

158. Sun, Z. *et al.* Giant nonreciprocal second-harmonic generation from antiferromagnetic bilayer CrI3. *Nature* **572**, 497–501 (2019).
159. Kirilyuk, A. & Rasing, T. Magnetization-induced-second-harmonic generation from surfaces and interfaces. *J. Opt. Soc. Am. B* **22**, 148–167 (2005).
160. Ni, Z. *et al.* Direct imaging of antiferromagnetic domains and anomalous layer-dependent mirror symmetry breaking in atomically thin MnPS_{3}. *Phys. Rev. Lett.* **127**, 187201 (2021).
161. Ni, Z. *et al.* Imaging the Néel vector switching in the monolayer antiferromagnet MnPSe3 with strain-controlled Ising order. *Nat. Nanotechnol.* **16**, 782–787 (2021).
162. Yariv, A. & Louisell, W. Theory of the optical parametric oscillator. *IEEE Journal of Quantum Electronics* **2**, 118–118 (1966).
163. Giordmaine, J. A. & Miller, R. C. Tunable coherent parametric oscillation in LiNbO3at optical frequencies. *Phys. Rev. Lett.* **14**, 973–976 (1965).
164. Giordmaine, J. A. Mixing of light beams in crystals. *Phys. Rev. Lett.* **8**, 19–20 (1962).
165. Canalias, C. & Pasiskevicius, V. Mirrorless optical parametric oscillator. *Nat. Photonics* **1**, 459–462 (2007).
166. Trovatello, C. *et al.* Optical parametric amplification by monolayer transition metal dichalcogenides. *Nat. Photonics* **15**, 6–10 (2021).
167. Yang, J. *et al.* Titanium:sapphire-on-insulator integrated lasers and amplifiers. *Nature* **630**, 853–859 (2024).
168. Zhang, Z. *et al.* Large-Scale Spectral Broadening of Femtosecond Optical Parametric Oscillators by MoTe 2 Films. *ACS Photonics* **11**, 1044–1050 (2024).
169. Wang, G. *et al.* Tunable nonlinear refractive index of two-dimensional MoS_2, WS_2, and MoSe_2 nanosheet dispersions [Invited]. *Photonics Res.* **3**, A51 (2015).
170. Catalano, J. Spontaneous parametric down-conversion and quantum entanglement. (Portland State University Library, 2017). doi:10.15760/honors.474.
171. Kwiat, P. G., Waks, E., White, A. G., Appelbaum, I. & Eberhard, P. H. Ultrabright source of polarization-entangled photons. *Phys. Rev. A* **60**, R773–R776 (1999).
172. Orieux, A., Versteegh, M. A. M., Jöns, K. D. & Ducci, S. Semiconductor devices for entangled photon pair generation: a review. *Rep. Prog. Phys.* **80**, 076001 (2017).
173. Li, X., Voss, P. L., Sharping, J. E. & Kumar, P. Optical-fiber source of polarization-entangled photons in the 1550 nm telecom band. *Phys. Rev. Lett.* **94**, 053601 (2005).
174. Kallioniemi, L. *et al.* Van der Waals engineering for quantum-entangled photon generation. *Nat. Photonics* 1–7 (2024).
175. Weissflog, M. A. *et al.* A tunable transition metal dichalcogenide entangled photon-pair source. *Nat. Commun.* **15**, 7600 (2024).
176. Feng, J. *et al.* Polarization-entangled photon-pair source with van der Waals 3R-WS2 crystal. *eLight* **4**, 16 (2024).
177. Li, L. *et al.* Metalens-array-based high-dimensional and multiphoton quantum source. *Science* **368**, 1487–1490 (2020).
178. Santiago-Cruz, T. *et al.* Resonant metasurfaces for generating complex quantum states. *Science* **377**, 991–995 (2022).
179. Hamel, D. R. *et al.* Direct generation of three-photon polarization entanglement. *Nat. Photonics* **8**, 801–807 (2014).
180. Zhang, Z. *et al.* High-performance quantum entanglement generation via cascaded second-order nonlinear processes. *Npj Quantum Inf.* **7**, 1–9 (2021).

181. Javid, U. A. *et al.* Ultrabroadband entangled photons on a nanophotonic chip. *Phys. Rev. Lett.* **127**, 183601 (2021).
182. Zograf, G. *et al.* Ultrathin 3R-$MoS_2$ metasurfaces with atomically precise edges for efficient nonlinear nanophotonics. *arXiv:2410.20960* (2024).
183. Hu, G. *et al.* Coherent steering of nonlinear chiral valley photons with a synthetic Au–WS2 metasurface. *Nat. Photonics* **13**, 467–472 (2019).
184. Wang, Z. *et al.* Two-dimensional materials for tunable and nonlinear metaoptics. *Adv. Photonics* **6**, (2024).
185. Jia, L. *et al.* Third-order optical nonlinearities of 2D materials at telecommunications wavelengths. *Micromachines (Basel)* **14**, 307 (2023).
186. Bristow, A. D., Rotenberg, N. & Van Driel, H. M. Two-photon absorption and Kerr coefficients of silicon for 850--2200nm. *Appl. Phys. Lett.* **90**, (2007).
187. Ikeda, K., Saperstein, R. E., Alic, N. & Fainman, Y. Thermal and Kerr nonlinear properties of plasma-deposited silicon nitride/ silicon dioxide waveguides. *Opt. Express* **16**, 12987 (2008).
188. Säynätjoki, A. *et al.* Ultra-strong nonlinear optical processes and trigonal warping in MoS2 layers. *Nat. Commun.* **8**, 893 (2017).
189. Wang, R. *et al.* Third-harmonic generation in ultrathin films of MoS2. *ACS Appl. Mater. Interfaces* **6**, 314–318 (2014).
190. Tang, T. *et al.* Third harmonic generation in thin NbOI2 and TaOI2. *Nanomaterials (Basel)* **14**, 412 (2024).
191. Soavi, G. *et al.* Broadband, electrically tunable third-harmonic generation in graphene. *Nat. Nanotechnol.* **13**, 583–588 (2018).
192. Jiang, T. *et al.* Gate-tunable third-order nonlinear optical response of massless Dirac fermions in graphene. *Nat. Photonics* **12**, 430–436 (2018).
193. Mak, K. F., Ju, L., Wang, F. & Heinz, T. F. Optical spectroscopy of graphene: From the far infrared to the ultraviolet. *Solid State Commun.* **152**, 1341–1349 (2012).
194. Youngblood, N., Peng, R., Nemilentsau, A., Low, T. & Li, M. Layer-tunable third-harmonic generation in multilayer black phosphorus. *ACS Photonics* **4**, 8–14 (2017).
195. Wang, X. & Lan, S. Optical properties of black phosphorus. *Adv. Opt. Photon., AOP* **8**, 618–655 (2016).
196. Wang, K. *et al.* Ultrafast nonlinear excitation dynamics of black phosphorus nanosheets from visible to mid-infrared. *ACS Nano* **10**, 6923–6932 (2016).
197. Siemens, M. E., Moody, G., Li, H., Bristow, A. D. & Cundiff, S. T. Resonance lineshapes in two-dimensional Fourier transform spectroscopy. *Opt. Express* **18**, 17699–17708 (2010).
198. Bristow, A. D. *et al.* A versatile ultrastable platform for optical multidimensional Fourier-transform spectroscopy. *Rev. Sci. Instrum.* **80**, 073108 (2009).
199. Cundiff, S. T. *et al.* Optical 2-D Fourier transform spectroscopy of excitons in semiconductor nanostructures. *IEEE J. Sel. Top. Quantum Electron.* **18**, 318–328 (2012).
200. Moody, G. *et al.* Intrinsic homogeneous linewidth and broadening mechanisms of excitons in monolayer transition metal dichalcogenides. *Nat. Commun.* **6**, 8315 (2015).
201. Cadiz, F. *et al.* Excitonic Linewidth Approaching the Homogeneous Limit in MoS2 -Based van der Waals Heterostructures. *Phys. Rev. X.* **7**, (2017).
202. Scuri, G. *et al.* Large excitonic reflectivity of monolayer MoSe_{2} encapsulated in hexagonal boron nitride. *Phys. Rev. Lett.* **120**, 037402 (2018).

203. Jakubczyk, T. *et al.* Radiatively limited dephasing and exciton dynamics in MoSe2 monolayers revealed with four-wave mixing microscopy. *Nano Lett.* **16**, 5333–5339 (2016).
204. Helmrich, S. *et al.* Phonon-assisted intervalley scattering determines ultrafast exciton dynamics in MoSe_{2} bilayers. *Phys. Rev. Lett.* **127**, 157403 (2021).
205. Huang, D. *et al.* Quantum dynamics of attractive and repulsive polarons in a doped MoSe2 monolayer. *Phys. Rev. X.* **13**, 011029 (2023).
206. Hao, K. *et al.* Direct measurement of exciton valley coherence in monolayer WSe2. *Nat. Phys.* **12**, 677–682 (2016).
207. Ma, C. *et al.* Recent progress in ultrafast lasers based on 2D materials as a saturable absorber. *Appl. Phys. Rev.* **6**, 041304 (2019).
208. Liu, W. *et al.* Recent Advances of 2D Materials in Nonlinear Photonics and Fiber Lasers. *Advanced Optical Materials* **8**, (2020).
209. Sun, Z. *et al.* Graphene mode-locked ultrafast laser. *ACS Nano* **4**, 803–810 (2010).
210. Wang, F., Dukovic, G., Brus, L. E. & Heinz, T. F. The optical resonances in carbon nanotubes arise from excitons. *Science* **308**, 838–841 (2005).
211. Ye, Z. *et al.* Probing excitonic dark states in single-layer tungsten disulphide. *Nature* **513**, 214–218 (2014).
212. Manca, M. *et al.* Enabling valley selective exciton scattering in monolayer WSe2 through upconversion. *Nat. Commun.* **8**, 14927 (2017).
213. Han, B. *et al.* Exciton states in monolayer MoSe2 and MoTe2 probed by upconversion spectroscopy. *Phys. Rev. X.* **8**, (2018).
214. Wang, Q. *et al.* High-energy gain upconversion in monolayer tungsten disulfide photodetectors. *Nano Lett.* **19**, 5595–5603 (2019).
215. Yu, Y. *et al.* Fundamental limits of exciton-exciton annihilation for light emission in transition metal dichalcogenide monolayers. *Phys. Rev. B.* **93**, (2016).
216. Ferray, M. *et al.* Multiple-harmonic conversion of 1064 nm radiation in rare gases. *J. Phys. B At. Mol. Opt. Phys.* **21**, L31–L35 (1988).
217. Ghimire, S. *et al.* Observation of high-order harmonic generation in a bulk crystal. *Nat. Phys.* **7**, 138–141 (2011).
218. Krause, J. L., Schafer, K. J. & Kulander, K. C. High-order harmonic generation from atoms and ions in the high intensity regime. *Phys. Rev. Lett.* **68**, 3535–3538 (1992).
219. Corkum, P. B. Plasma perspective on strong field multiphoton ionization. *Phys. Rev. Lett.* **71**, 1994–1997 (1993).
220. Corkum, P. B. & Krausz, F. Attosecond science. *Nat. Phys.* **3**, 381–387 (2007).
221. Vampa, G. *et al.* All-optical reconstruction of crystal band structure. *Phys. Rev. Lett.* **115**, 193603 (2015).
222. Goulielmakis, E. & Brabec, T. High harmonic generation in condensed matter. *Nat. Photonics* **16**, 411–421 (2022).
223. Liu, H. *et al.* High-harmonic generation from an atomically thin semiconductor. *Nat. Phys.* **13**, 262–265 (2017).
224. Heide, C. *et al.* High-harmonic generation from artificially stacked 2D crystals. *Nanophotonics* **12**, 255–261 (2023).
225. Maroju, P. K. *et al.* Attosecond pulse shaping using a seeded free-electron laser. *Nature* **578**, 386–391 (2020).
226. Franken, P. A. & Ward, J. F. Optical harmonics and nonlinear phenomena. *Rev. Mod. Phys.* **35**, 23–39 (1963).

227. Chang, D. E., Sørensen, A. S., Demler, E. A. & Lukin, M. D. A single-photon transistor using nanoscale surface plasmons. *Nat. Phys.* **3**, 807–812 (2007).
228. Hwang, J. *et al.* A single-molecule optical transistor. *Nature* **460**, 76–80 (2009).
229. Turchette, Q. A., Hood, C. J., Lange, W., Mabuchi, H. & Kimble, H. J. Measurement of conditional phase shifts for quantum logic. *Phys. Rev. Lett.* **75**, 4710–4713 (1995).
230. Volz, T. *et al.* Ultrafast all-optical switching by single photons. *Nat. Photonics* **6**, 605–609 (2012).
231. Schuster, I. *et al.* Nonlinear spectroscopy of photons bound to one atom. *Nat. Phys.* **4**, 382–385 (2008).
232. Lounis, B. & Moerner, W. E. Single photons on demand from a single molecule at room temperature. *Nature* **407**, 491–493 (2000).
233. Castelletto, S. *et al.* A silicon carbide room-temperature single-photon source. *Nat. Mater.* **13**, 151–156 (2014).
234. Harris, S. E., Field, J. E. & Imamoglu, A. Nonlinear optical processes using electromagnetically induced transparency. *Phys. Rev. Lett.* **64**, 1107–1110 (1990).
235. Fushman, I. *et al.* Controlled phase shifts with a single quantum dot. *Science* **320**, 769–772 (2008).
236. Shahnazaryan, V., Iorsh, I., Shelykh, I. A. & Kyriienko, O. Exciton-exciton interaction in transition-metal dichalcogenide monolayers. *Phys. Rev. B.* **96**, 115409 (2017).
237. Tagarelli, F. *et al.* Electrical control of hybrid exciton transport in a van der Waals heterostructure. *Nat. Photonics* **17**, 615–621 (2023).
238. Zhang, L. *et al.* Electrical control and transport of tightly bound interlayer excitons in a MoSe_{2}/hBN/MoSe_{2} heterostructure. *Phys. Rev. Lett.* **132**, 216903 (2024).
239. Rapaport, B. L. A. Exciton correlations in coupled quantum wells and their luminescence blue shift. *Physics Review B* **80**, 195313 (2009).
240. Movva, H. C. P. *et al.* Tunable Γ-K valley populations in hole-doped trilayer WSe_{2}. *Phys. Rev. Lett.* **120**, 107703 (2018).
241. Uto, T. *et al.* Interaction-Induced ac Stark Shift of Exciton-Polaron Resonances. *Phys. Rev. Lett.* **132**, 056901 (2024).
242. Thureja, D. *et al.* Electrically defined quantum dots for bosonic excitons. *arXiv:2402.19278* (2024).
243. Tang, Y. *et al.* Tuning layer-hybridized moiré excitons by the quantum-confined Stark effect. *Nat. Nanotechnol.* **16**, 52–57 (2021).
244. Brotons-Gisbert, M. *et al.* Spin–layer locking of interlayer excitons trapped in moiré potentials. *Nat. Mater.* **19**, 630–636 (2020).
245. Kim, D. S. *et al.* An active metasurface enhanced with moiré ferroelectricity. *arXiv:2405.11159* (2024).
246. Liuxin, G. *et al.* Quantum confining excitons with electrostatic moiré superlattice. *arXiv:2501.11713* (2025).
247. Li, W., Lu, X., Dubey, S., Devenica, L. & Srivastava, A. Dipolar interactions between localized interlayer excitons in van der Waals heterostructures. *Nat. Mater.* **19**, 624–629 (2020).
248. Sie, E. J. *et al.* Valley-selective optical Stark effect in monolayer WS2. *Nat. Mater.* **14**, 290–294 (2015).
249. Cunningham, P. D., Hanbicki, A. T., Reinecke, T. L., McCreary, K. M. & Jonker, B. T. Resonant optical Stark effect in monolayer WS2. *Nat. Commun.* **10**, 5539 (2019).

250. Schmitt-Rink, S., Chemla, D. S. & Miller, D. A. B. Linear and nonlinear optical properties of semiconductor quantum wells. *Adv. Phys.* **38**, 89–188 (1989).
251. Kobayashi, Y. *et al.* Floquet engineering of strongly driven excitons in monolayer tungsten disulfide. *Nat. Phys.* **19**, 171–176 (2023).
252. Yong, C.-K. *et al.* Valley-dependent exciton fine structure and Autler-Townes doublets from Berry phases in monolayer MoSe2. *Nat. Mater.* **18**, 1065–1070 (2019).
253. Zhou, L. *et al.* Cavity Floquet engineering. *Nat. Commun.* **15**, 7782 (2024).
254. Sarkar, S. *et al.* Sub-wavelength optical lattice in 2D materials. *arXiv:2406.00464* (2024).
255. Kim, H., Dehghani, H., Aoki, H., Martin, I. & Hafezi, M. Optical imprinting of superlattices in two-dimensional materials. *Phys. Rev. Res.* **2**, (2020).
256. Deng, H., Haug, H. & Yamamoto, Y. Exciton-polariton Bose-Einstein condensation. *Rev. Mod. Phys.* **82**, 1489–1537 (2010).
257. Saba, M. Intrinsic non-linearities in exciton-cavity-coupled systems. *Physica B Condens. Matter* **272**, 472–475 (1999).
258. Ciuti, C., Schwendimann, P. & Quattropani, A. Theory of polariton parametric interactions in semiconductor microcavities. *Semicond. Sci. Technol.* **18**, S279–S293 (2003).
259. G. Rochat, C. Ciuti, V. Savona, C. Piermarocchi, A. Quattropani, and P. Schwendimann. Excitonic Bloch equations for a two-dimensional system of interacting excitons. *Phys. Rev. B.* **61**, 13856–13862 (2000).
260. Shahnazaryan, V., Kozin, V. K., Shelykh, I. A., Iorsh, I. V. & Kyriienko, O. Tunable optical nonlinearity for transition metal dichalcogenide polaritons dressed by a Fermi sea. *Phys. Rev. B.* **102**, (2020).
261. Stepanov, P. *et al.* Exciton-exciton interaction beyond the hydrogenic picture in a MoSe_{2} monolayer in the strong light-matter coupling regime. *Phys. Rev. Lett.* **126**, 167401 (2021).
262. Wei, K. *et al.* Charged biexciton polaritons sustaining strong nonlinearity in 2D semiconductor-based nanocavities. *Nat. Commun.* **14**, 5310 (2023).
263. Emmanuele, R. P. A. *et al.* Highly nonlinear trion-polaritons in a monolayer semiconductor. *Nat. Commun.* **11**, 3589 (2020).
264. Louca, C. *et al.* Interspecies exciton interactions lead to enhanced nonlinearity of dipolar excitons and polaritons in MoS2 homobilayers. *Nat. Commun.* **14**, 3818 (2023).
265. Wild, D. S., Shahmoon, E., Yelin, S. F. & Lukin, M. D. Quantum nonlinear optics in atomically thin materials. *Phys. Rev. Lett.* **121**, 123606 (2018).
266. Zhang, L. *et al.* Van der Waals heterostructure polaritons with moiré-induced nonlinearity. *Nature* **591**, 61–65 (2021).
267. Park, H. *et al.* Dipole ladders with large Hubbard interaction in a moiré exciton lattice. *Nat. Phys.* **19**, 1286–1292 (2023).
268. Xiong, R. *et al.* Correlated insulator of excitons in WSe2/WS2 moiré superlattices. *Science* **380**, 860–864 (2023).
269. Gao, B. *et al.* Excitonic Mott insulator in a Bose-Fermi-Hubbard system of moiré WS2/WSe2 heterobilayer. *Nat. Commun.* **15**, 2305 (2024).
270. A. Verger, C. Ciuti, and I. Carusotto. Polariton quantum blockade in a photonic dot. *Phys. Rev. B Condens. Matter Mater. Phys.* **73**, 193306 (2006).
271. Muñoz-Matutano, G. *et al.* Emergence of quantum correlations from interacting fibre-cavity polaritons. *Nature materials* vol. 18 213–218 (2019).
272. Somaschi, N. *et al.* Near-optimal single-photon sources in the solid state. *Nat. Photonics* **10**, 340–345 (2016).

273. He, Y.-M. *et al.* On-demand semiconductor single-photon source with near-unity indistinguishability. *Nat. Nanotechnol.* **8**, 213–217 (2013).
274. Delteil, A. *et al.* Towards polariton blockade of confined exciton-polaritons. *Nat. Mater.* **18**, 219–222 (2019).
275. Moody, G., Chang, L., Steiner, T. J. & Bowers, J. E. Chip-scale nonlinear photonics for quantum light generation. *AVS Quantum Sci.* **2**, 041702 (2020).
276. Elshaari, A. W., Pernice, W., Srinivasan, K., Benson, O. & Zwiller, V. Hybrid integrated quantum photonic circuits. *Nat. Photonics* **14**, 285–298 (2020).
277. Sirleto, L. & Righini, G. An introduction to nonlinear integrated photonics: Structures and devices. *Micromachines* **14**, 614 (2023).
278. Bogdanov, A. A., Makarov, S. & Kivshar, Y. New frontiers in nonlinear nanophotonics. *Nanophotonics* **13**, 3175–3179 (2024).
279. Perea-Causín, R. *et al.* Exciton optics, dynamics, and transport in atomically thin semiconductors. *APL Mater.* **10**, 100701 (2022).
280. Li, W., Brumme, T. & Heine, T. Relaxation effects in transition metal dichalcogenide bilayer heterostructures. *Npj 2D Mater. Appl.* **8**, 1–11 (2024).
281. Mannix, A. J. *et al.* Robotic four-dimensional pixel assembly of van der Waals solids. *Nat. Nanotechnol.* **17**, 361–366 (2022).
282. Huang, D., Choi, J., Shih, C.-K. & Li, X. Excitons in semiconductor moiré superlattices. *Nat. Nanotechnol.* **17**, 227–238 (2022).
283. Yang, X., Wang, X., Faizan, M., He, X. & Zhang, L. Second-harmonic generation in 2D moiré superlattices composed of bilayer transition metal dichalcogenides. *Nanoscale* **16**, 2913–2922 (2024).
284. Camacho-Guardian, A. & Cooper, N. R. Moiré-induced optical nonlinearities: Single- and multiphoton resonances. *Phys. Rev. Lett.* **128**, 207401 (2022).
285. Mennel, L., Paur, M. & Mueller, T. Second harmonic generation in strained transition metal dichalcogenide monolayers: MoS2, MoSe2, WS2, and WSe2. *APL Photonics* **4**, 034404 (2019).
286. Khestanova, E. *et al.* Electrostatic control of nonlinear photonic-crystal polaritons in a monolayer semiconductor. *Nano Lett.* **24**, 7350–7357 (2024).
287. Moody, G., Schaibley, J. & Xu, X. Exciton dynamics in monolayer transition metal dichalcogenides. *J. Opt. Soc. Am. B* **33**, C39–C49 (2016).
288. Esmaeil Zadeh, I. *et al.* Efficient Single-Photon Detection with 7.7 ps Time Resolution for Photon-Correlation Measurements. *ACS Photonics* **7**, 1780–1787 (2020).
289. Korzh, B. *et al.* Demonstration of sub-3 ps temporal resolution with a superconducting nanowire single-photon detector. *Nat. Photonics* **14**, 250–255 (2020).
290. Choudhury, T. H., Zhang, X., Al Balushi, Z. Y., Chubarov, M. & Redwing, J. M. Epitaxial Growth of Two-Dimensional Layered Transition Metal Dichalcogenides. *Annu. Rev. Mater. Res.* **50**, 155–177 (2020).
291. Zhang, T., Wang, J., Wu, P., Lu, A.-Y. & Kong, J. Vapour-phase deposition of two-dimensional layered chalcogenides. *Nat. Rev. Mater.* **8**, 799–821 (2023).
292. Kim, J.-Y., Ju, X., Ang, K.-W. & Chi, D. Van der Waals layer transfer of 2D materials for monolithic 3D electronic system integration: Review and outlook. *ACS Nano* **17**, 1831–1844 (2023).
293. Li, H. *et al.* Imaging moiré flat bands in three-dimensional reconstructed WSe2/WS2 superlattices. *Nat. Mater.* **20**, 945–950 (2021).

294. Zhang, Y. *et al.* Atom-by-atom imaging of moiré transformations in 2D transition metal dichalcogenides. *Sci. Adv.* **10**, eadk1874 (2024).
295. Frisenda, R. *et al.* Recent progress in the assembly of nanodevices and van der Waals heterostructures by deterministic placement of 2D materials. *Chem. Soc. Rev.* **47**, 53–68 (2018).
296. Le, C. T. *et al.* Nonlinear optical characteristics of monolayer MoSe2. *Ann. Phys.* **528**, 551–559 (2016).
297. Chen, H. *et al.* Enhanced second-harmonic generation from two-dimensional MoSe2 on a silicon waveguide. *Light Sci. Appl.* **6**, e17060 (2017).
298. Autere, A. *et al.* Optical harmonic generation in monolayer group-VI transition metal dichalcogenides. *Phys. Rev. B Condens. Matter* **98**, (2018).
299. Janisch, C. *et al.* Extraordinary Second Harmonic Generation in tungsten disulfide monolayers. *Sci. Rep.* **4**, 5530 (2014).
300. Ribeiro-Soares, J. *et al.* Second Harmonic Generation in WSe2. *2D Materials* **2**, 045015 (2015).
301. Rosa, H. G. *et al.* Characterization of the second- and third-harmonic optical susceptibilities of atomically thin tungsten diselenide. *Sci. Rep.* **8**, 10035 (2018).
302. Kim, S. *et al.* Second-harmonic generation in multilayer hexagonal boron nitride flakes. *Opt. Lett.* **44**, 5792–5795 (2019).

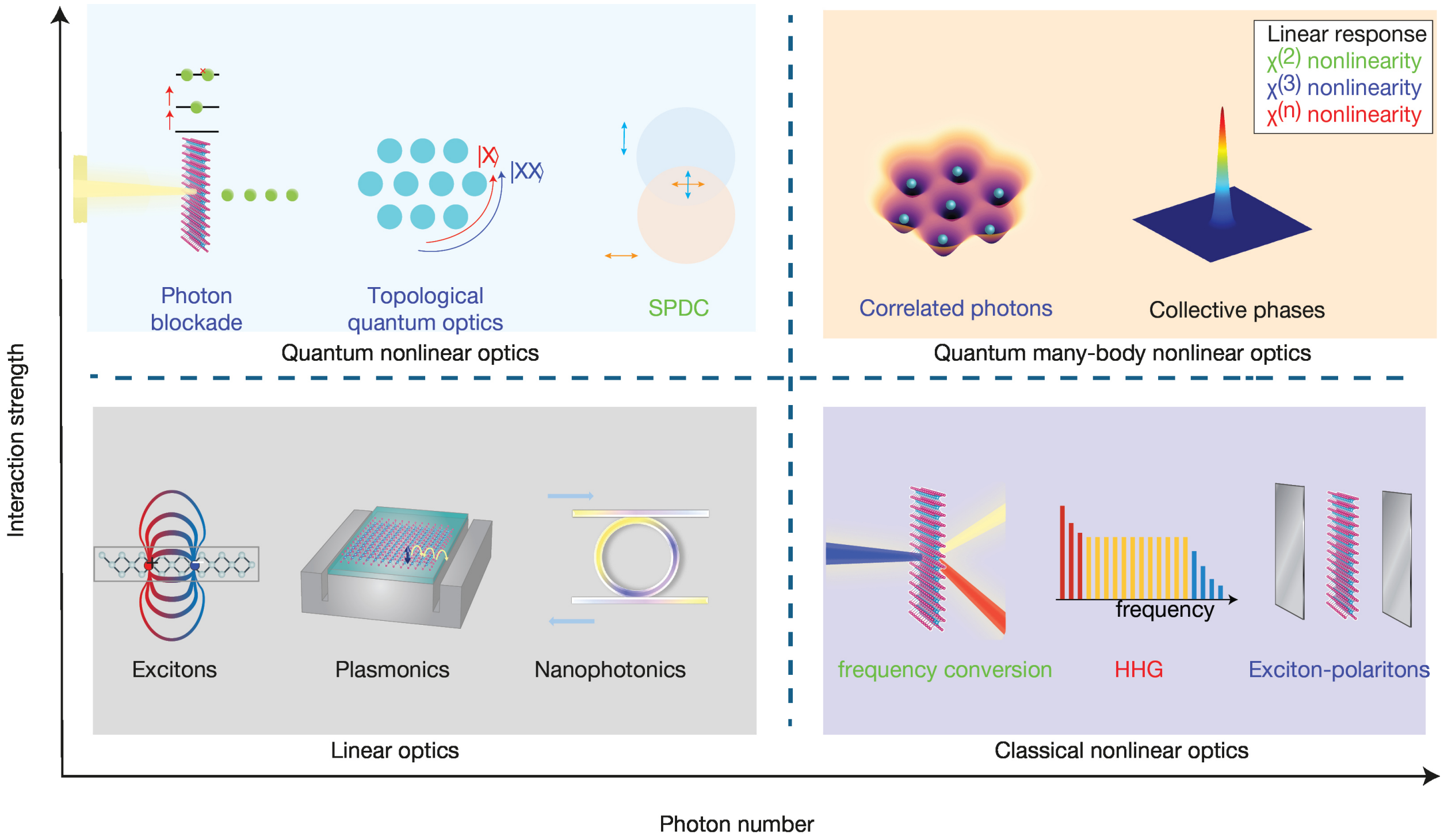

**Figure 1. Nonlinear optical phenomena in different regimes.** We can classify the many emerging optical phenomena in terms of the interaction strength among photons and the photon number (light intensity). The lower left vertical panel represents the classical linear optics regime, with a low number of optical excitations that are weakly interacting. Examples include excitons, plasmons, and nanophotonic devices. With an increasing number of photons, the nonlinear optical response becomes strong, leading to classical nonlinear optical phenomena, such as frequency conversion. Exciton-polaritons form when embedding excitons inside a cavity, which enhances their interactions and features intriguing properties such as condensates at high photon numbers. An interesting regime emerges in the upper left corner, where the interaction strength is so large that nonlinear effects become obvious at the few-photon regime. Examples of such include photon blockade, topological quantum systems, and entangled photon pair generation via spontaneous parametric down-conversion (SPDC). Finally, for strong interactions and large photon numbers, the system behaves as quantum many-body systems with emergent properties, including correlated Bosonic lattices and quantum fluids, which remain largely unexplored.

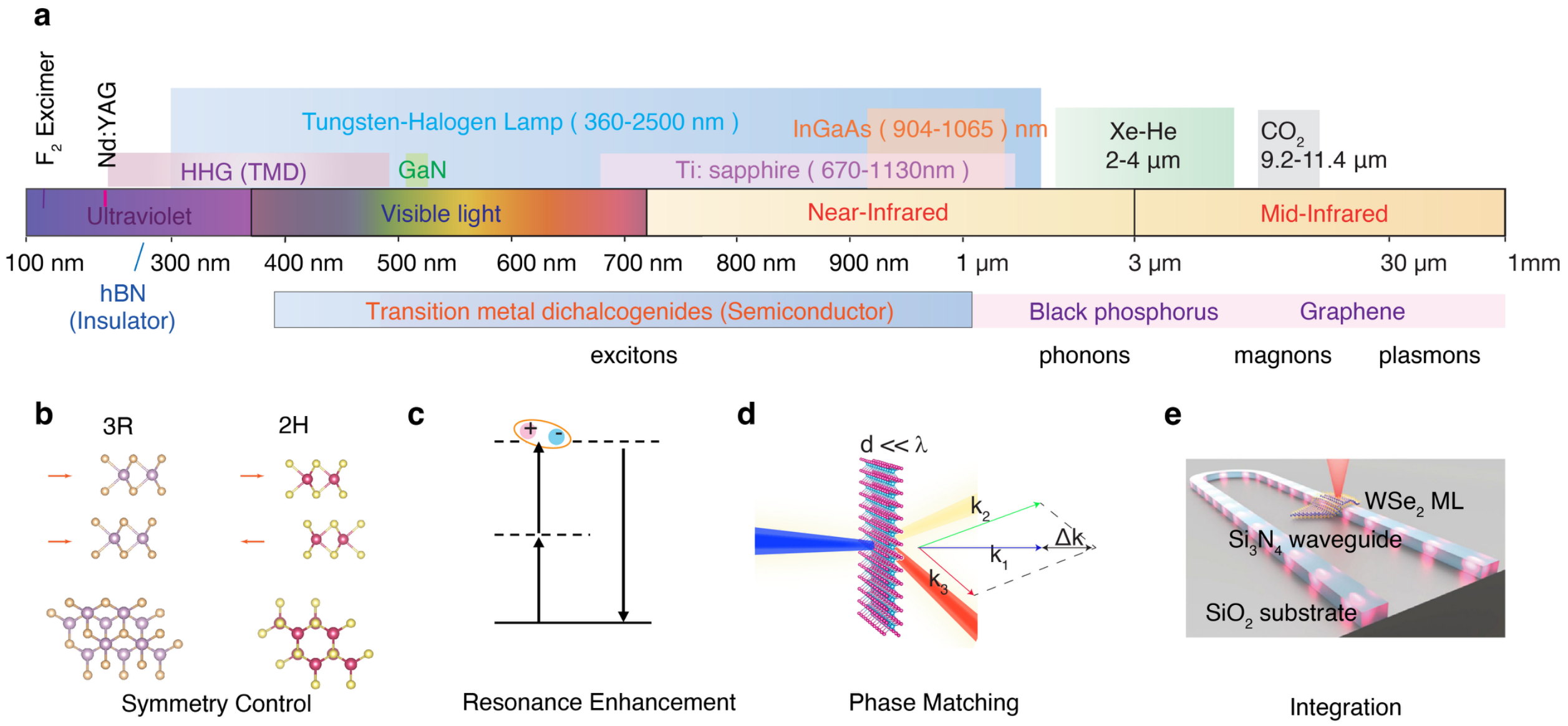

**Figure 2**. **Key features of 2D materials relevant to nonlinear optics.** **a,** Electromagnetic spectrum from UV to MIR with the common light source. 2D materials offer distinct electronic and optical properties, covering a wide spectral range. They also feature various optical excitations, such as excitons and magnons, which can be used to enhance nonlinear response resonantly. **b,** Control of lattice symmetry in 2D heterostructures by controlling the stacking between different layers. **c,** Enhancement of nonlinear optical response via resonance such as excitons. **d,** Relaxed phase matching conditions due to small propagation length. **e**, Integrating 2D materials into photonic structures such as waveguides. **e**, Reproduced with permission from *ACS Photonics* **8**, 1069–1076 (2021). Copyright 2021 American Chemical Society.

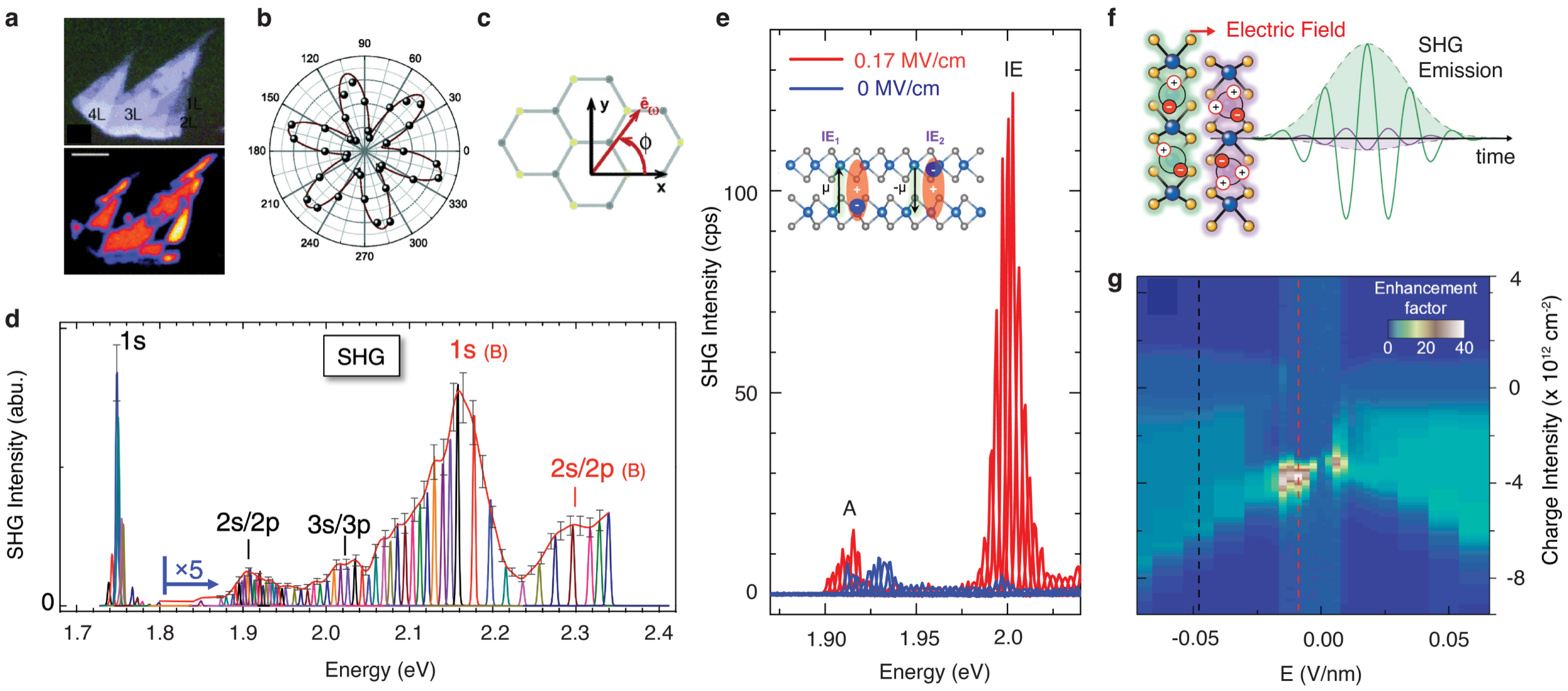

**Figure 3. Symmetry control, resonance enhancement, and electrical tuning of the second-order nonlinearity. a,** Top panel: optical image of multilayer 2H $MoS_2$ on quartz substrate. Bottom panel: SHG image for the same flake under pumping wavelength of 860nm. Due to the inversion symmetry, odd and even layers exhibit different SHG intensity contrasts. **b,** Monolayer $MoS_2$ polar plot of the second-harmonic intensity as a function of the sample angle. **c**, top view of the $MoS_2$ crystallographic orientation. The red arrow refers to the excitation laser polarization. **d**, Monolayer $WSe_2$ SHG spectroscopy at T = 4 K. SHG signal exhibits significant enhancement when the $2\hbar\omega$ is resonant at the exciton frequency. **e**, SHG spectra of A and interlayer exciton (IE) resonance at $F_Z = 0$ MV/cm and $F_Z = 0.17$ MV/cm in homobilayer $MoS_2$. The inset is the schematic of the interlayer excitons in bilayer $MoS_2$. The SHG signal at $F_Z = 0.17$ MV/cm increases by a factor of 25 compared to that at zero electric field. This enhancement is due to the Stark shift of the IE with the static field. At $F_Z = 0.17$ MV/cm, IE splits into two energy $IE_1$ and $IE_2$, which both contribute to the SHG signal. This enables tuning SHG signal of interlayer exciton (IE) in $MoS_2$ bilayer via applied electric field. **f**, Schematic of the broken symmetry in bilayer $WSe_2$ under certain doping with electric field induce SHG signal enhancement. **g**, SHG enhancement 2D map with changes of electric field and doping level. Under certain electric fileds and doping, the SHG enhancement reaches the maximum factor of 40. **a-c,** Reproduced with permission from *Phys. Rev. B.* **87**, 201401 (2013). Copyright 2013 American Physical Society. **d**, Reproduced with permission from *Phys. Rev. Lett*. **114**, 097403 (2015). Copyright 2015 American Physical Society. **e**, Reproduced with permission from *Nat. Commun*. **12**, 6894 (2021). Copyright 2021 Springer Nature Ltd. **f-g**, Reproduced with permission from *Nano Lett*. **24**, 14847–14853 (2024). Copyright 2024 American Chemical Society.

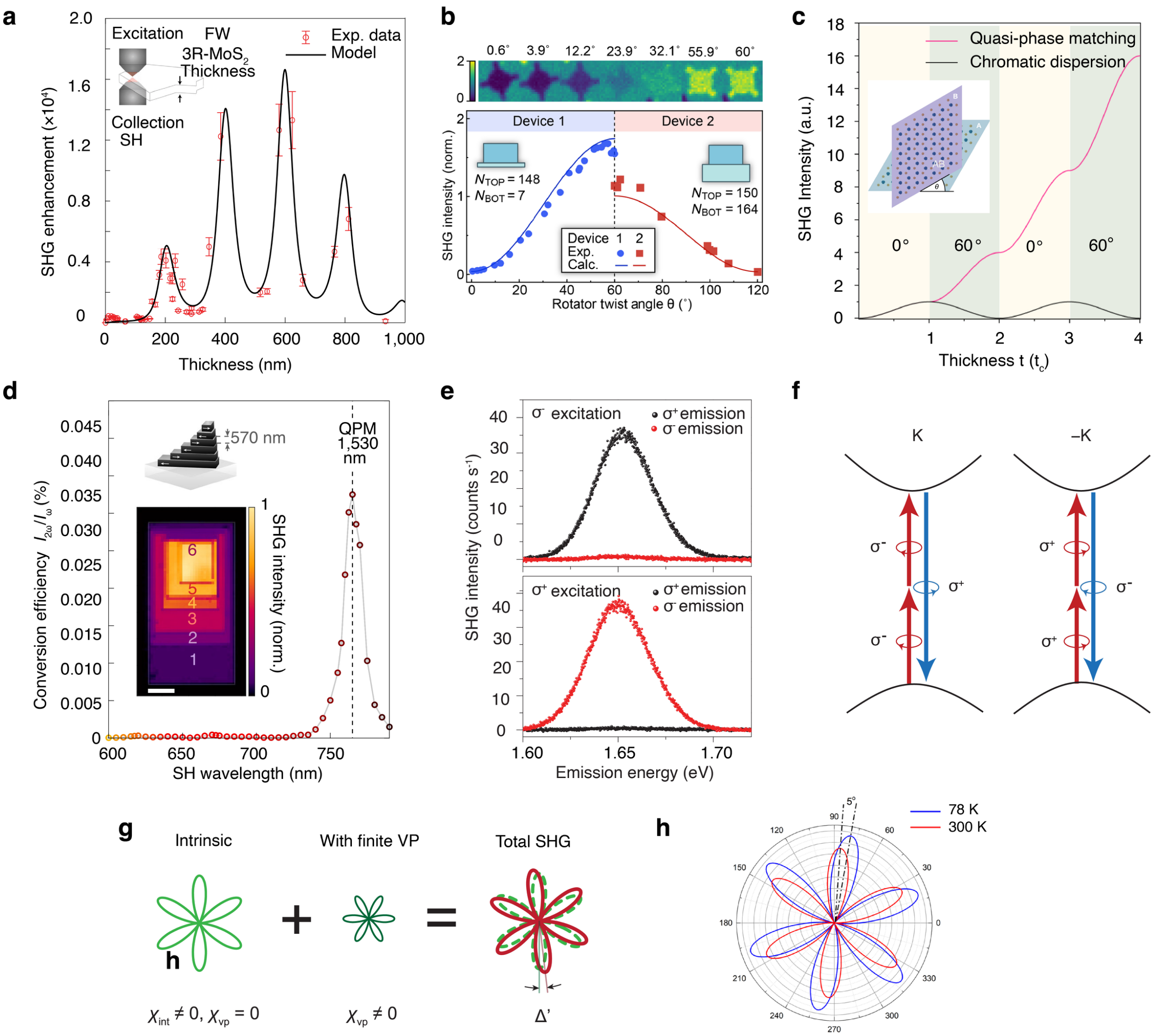

**Figure 4. Polarization dependence and phase matching of second order nonlinear processes in 2D materials. a**, Measured SHG enhancement of the 3R-$MoS_2$ compared to the monolayer $MoS_2$ with the change of thickness. Under the pump photon energy of 0.815 eV, the coherence length $L_c$ is around 530nm, with transmittance period of 182 nm. **b**, SHG intensity changes as the twist angle from 0º (AA') to 60º (AB) between two hBN flakes. The giant enhancement of the SHG intensity at the AB interface is due to the broken symmetry. **c**, Simulated SHG intensity change as a function of the total thickness of the twisted 3R-$MoS_2$ with a twist angle of 60º. Inset is the schematic of the stacking of the two 3R-$MoS_2$ flakes with the thickness of the coherent length $t_c$. For the quasi-phase-matching (QPM) scenario (pink line), periodic twisting enables quadratic growth of the SHG field with increasing thickness. In contrast, without QPM (black line), chromatic dispersion induces destructive interference once the thickness exceeds $t_c$, limiting the overall SHG intensity. **d**, SHG conversion efficiency ($I_{2\omega}/I_{\omega}$) in stacked PPTMD(3R-$MoS_2$) with pump wavelength of 1530nm. Each layer in the PPTMD has a thickness of 570 nm. **e,** Circular polarization resolved SHG spectra show that the SHG from monolayer $WSe_2$ is cross polarized. The angular momentum mismatch between the absorbed and emitted photons is provided by the lattice with broken inversion symmetry. **f**, Schematic of the interband valley optical selection rule. **g**, Schematic of the valley polarization induced non-zero second nonlinear susceptibility modified the intrinsic SHG polarization with a change of the angle relative to the main crystallographic axis. **h**, Fitted polarization-resolved SHG measured at different temperatures. Changes in polarization angle and intensity indicate variations in valley imbalance. Lower temperatures preserve valley imbalance, resulting in higher SHG intensity. **a,** Reproduced with permission from *Nat. Photon*. **16**, 698–706 (2022). Copyright 2022 Springer Nature Ltd. **b**, Reproduced with permission from *Sci. Adv*. **7**, eabe8691 (2021). Copyright 2021 American Association for the Advancement of Science. **c,** Reproduced with permission from *Nat. Commun*. **15**, 9979 (2024). Copyright 2024 Springer Nature Ltd. **d,** Reproduced with permission from *Nat. Photon*. 1–9 (2025). Copyright 2025 Springer Nature Ltd. **e, f,** Reproduced with permission from *Nat. Nanotechnol*. **10**, 407–411 (2015). Copyright 2015 Springer Nature Ltd. **g**, Reproduced with permission from *ACS Photonics.* **7**, 925–931 (2020). Copyright 2020 American Chemical Society. **h**, Reproduced with permission from *Npj 2D Mater. Appl*. **5**, 1–9 (2021). Copyright 2021 Springer Nature Ltd.

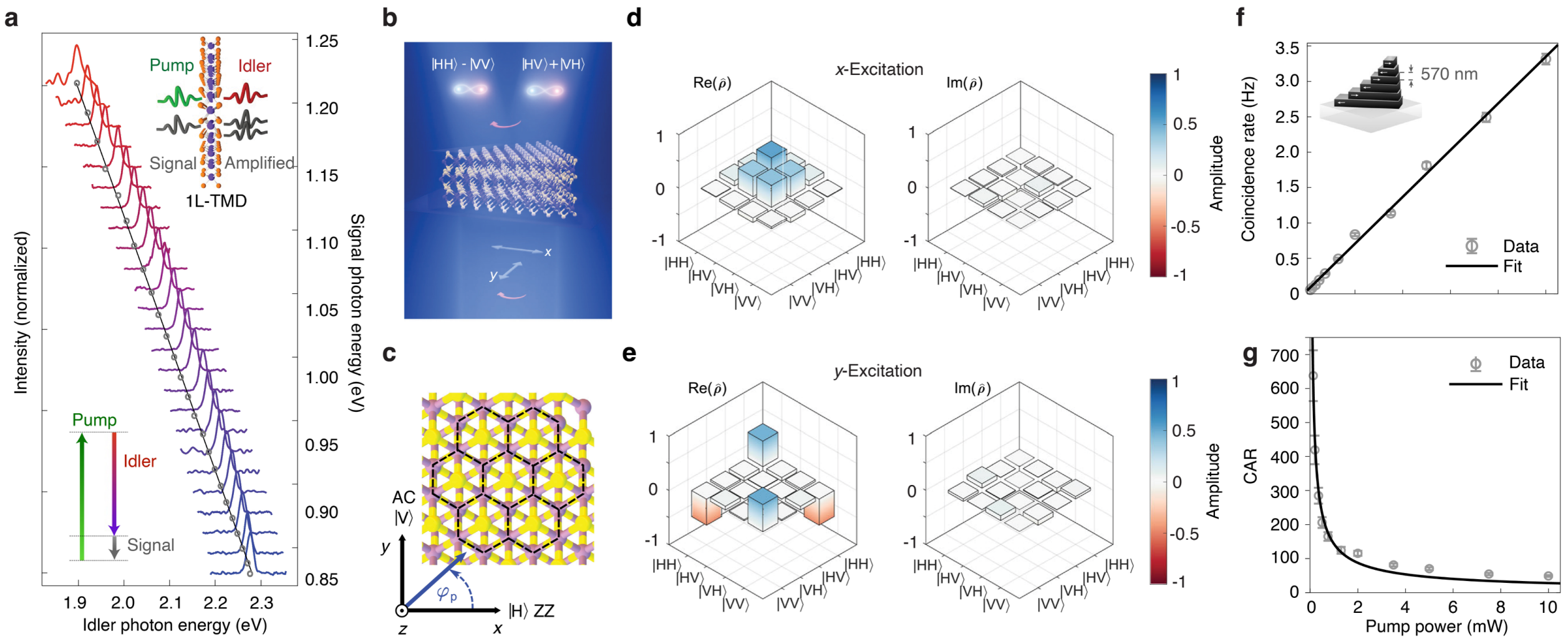

**Figure 5. Parametric amplification and down-conversion. a,** Monolayer TMD as broadband tunable OPA: the normalized tunable idler spectra measured on monolayer MoSe2 in a broad photon energy range from 1.90 eV to 2.28 eV. The spectral window is limited by the tunability of the signal beam used in the paper. **b-e, Generation of entangled photon pairs through SPDC**. **b**, Schematic of generation pairs of polarization-entangled signal and idler photons via SPDC process in a 3R-stacked $MoS_2$. Different maximally entangled Bell states are generated by changing excitation polarization. **c**, Top view of the 3R $MoS_2$ stack crystalline structure. The x and y directions (also corresponding to |H> and |V> polarization direction) are aligned with zigzag (ZZ) and armchair (AC) directions in the crystal structure. The blue arrow defines the pump polarization with the angle $\varphi_p$. **d, e**, Measured density matrix $\hat{\rho}$ of polarization quantum state for x and y pump direction. For both x and y linear polarizations, different maximally entangled states can be generated. **f, g**, Coincidence characterization of the generated photon pairs in periodically poled TMD (PPTMD) with six periods. The coincidence-to-accidental ratio (CAR), is defined as $CAR = \frac{R_{SPDC}}{R_C}$, with $R_{SPDC}$ refers to SPDC rate and is corrected as $R_{SPDC} = R_C - R_{acc}$. $R_{acc}$ is the accidental coincidence count rate. CAR in this frame reaches a maximum of $638 \pm 35$ at the telecom wavelength. **a**, Reproduced with permission from *Nat. Photon*. **15**, 6–10 (2021). Copyright 2021 Springer Nature Ltd. **b-e**, Reproduced with permission from *Nat. Commun*. **15**, 7600 (2024). Copyright 2024 Springer Nature Ltd. **f-g**, Reproduced with permission from *Nat. Photon.* 1–9 (2025). Copyright 2025 Springer Nature Ltd.

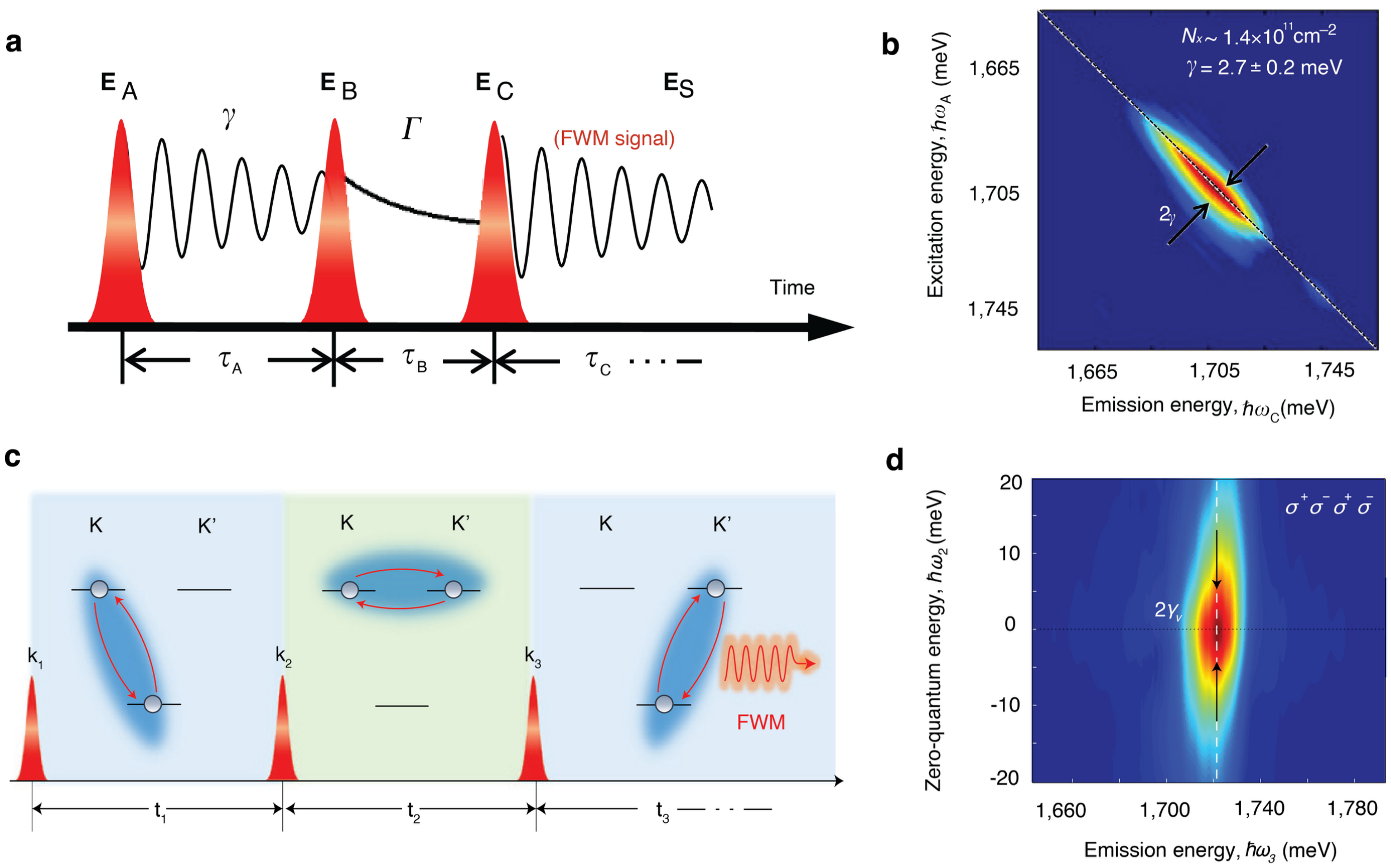

**Figure 6. Two-dimensional coherent spectroscopy utilizing four-wave mixing. a,** Schematic of the 2D coherent spectroscopy technique. Three phase-stabilized pulses radiate on the sample and then emit the FWM signal with the sequence shown in the figure. Scanning $\tau_A$ with $\tau_B$ fixed explore the coherent nature ($\gamma$) for excitons in TMD, while scan $\tau_B$ with $\tau_A$ fixed explore the population relaxation $\Gamma$.**b**, 2D Fourier-transform spectra of the echo signal revealing the linewidth broadening mechanism. The intrinsic homogeneous linewidth of the exciton resonance corresponds to the cross-diagonal line (dashed line), while the total linewidth includes inhomogeneous broadening. The exciton dephasing rate $\gamma$ can be extrapolated from the linewidth. **c-d**, Detection of valley coherence in monolayer TMD. **c,** As the first and third pulses are co-polarized ($\sigma^+$) and the second and detected polarization are co-polarized but with opposite helicity ($\sigma$-), the first pulse ($\sigma^+$)creates exciton population in K valley with a decay rate of $\Gamma_k$. The second pulse ($\sigma$-) creates a coherent superposition between K and K' valleys with a coherent decay rate of $\gamma_v$. The third pulse then converts the non-radiative valley coherence to optical coherence in K' valley between the exciton and ground state. **d,** 2D spectrum using alternative helicity of the excitation and collection pulses. As delay $t_1$ is fixed and $t_2$ is swept, the zero-quantum energy $\hbar\omega_2$ is given during $t_2$ by Fourier transform. The width of the line shape along $\hbar\omega_3$ demonstrate the measured valley coherence decay rate of $\gamma_v = 6.9 \pm 0.2$ meV. **a,b**, Reproduced with permission from *Nat. Commun*. **6**, 8315 (2015). Copyright 2015 Springer Nature Ltd. **c,d,** Reproduced with permission from *Nat. Phys.* **12**, 677–682 (2016). Copyright 2016 Springer Nature Ltd.

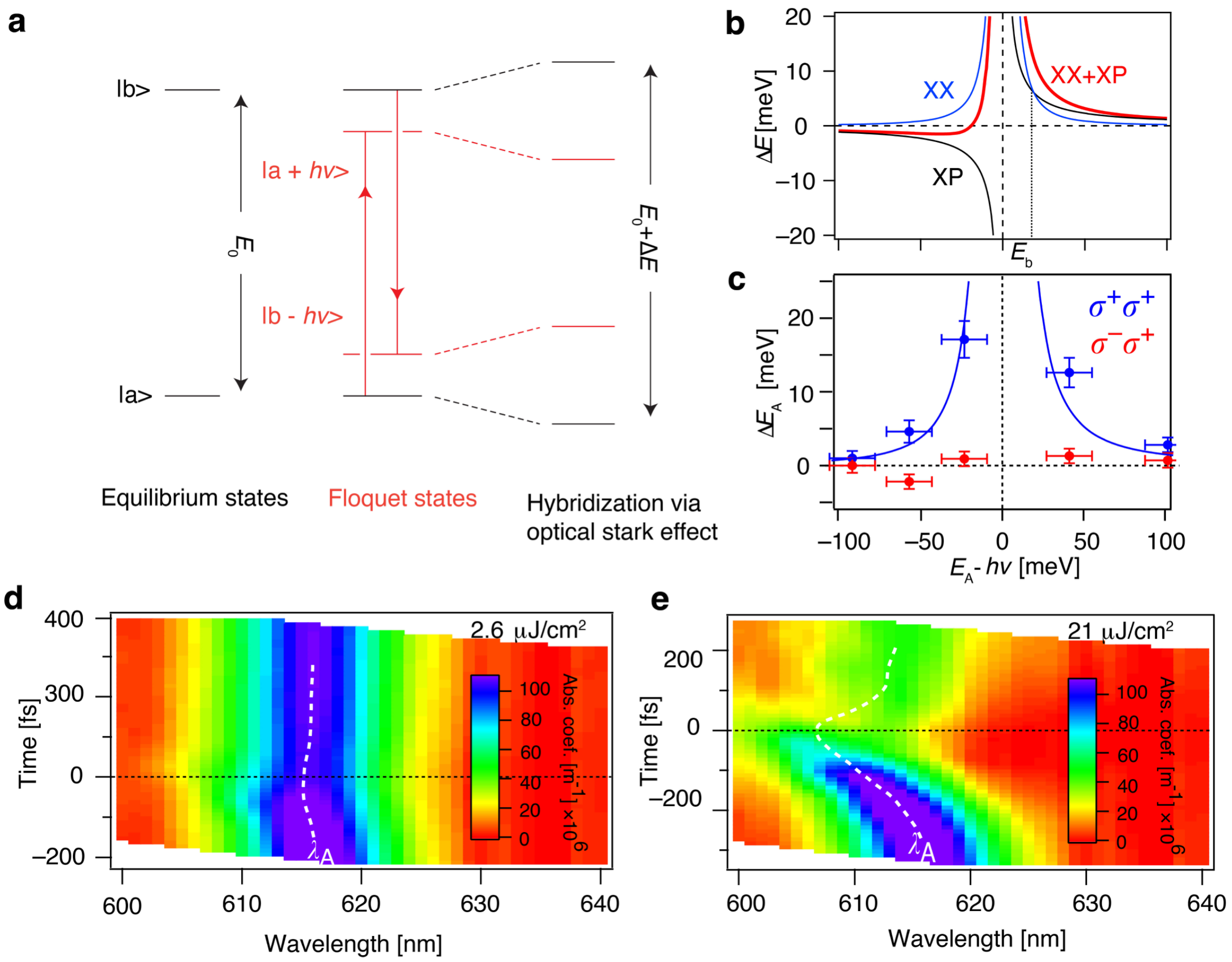

**Figure 7. Valley selective optical Stark effect in monolayer TMD. a,** Energy level diagram of two-level |a> and |b> showing that with applied light field, coherent absorption of light by |a> form the photon-dressed state $|a + hv>$ and $|b - hv>$, these dressed states hybridize with the equilibrium state through the electric field and finally result in the energy splitting. **b**, Excitonic model of the optical Stark effect. The black curve (XP) refers to the exciton-photon contribution, blue curve (XX) refers to the exciton-exciton contribution. The red curve refers to the sum of the two interactions. The gray curve refers to the exciton binding energy used in the model, which is 20 meV. **c**, Measured valley-selective optical Stark effect near A exciton resonance in monolayer $WS_2$. The energy blue shift in the co-polarized case under the above-band excitation is due to the repulsion among virtual excitons. **d, e,** Power dependence of the optical Stark effect shown by the time-dependent absorption spectra. Co-circular pulse with excitation wavelength of 610 nm is used as pump. Under the largest pump power in **e**, optical Stark shift of 32± 5 meV is observed. **a**, Reproduced with permission from *Nat. Mater.* **14**, 290–294 (2015). Copyright 2015 Springer Nature Ltd. **b-e**, Reproduced with permission from *Nat. Commun.* **10**, 5539 (2019). Copyright 2019 Springer Nature Ltd.

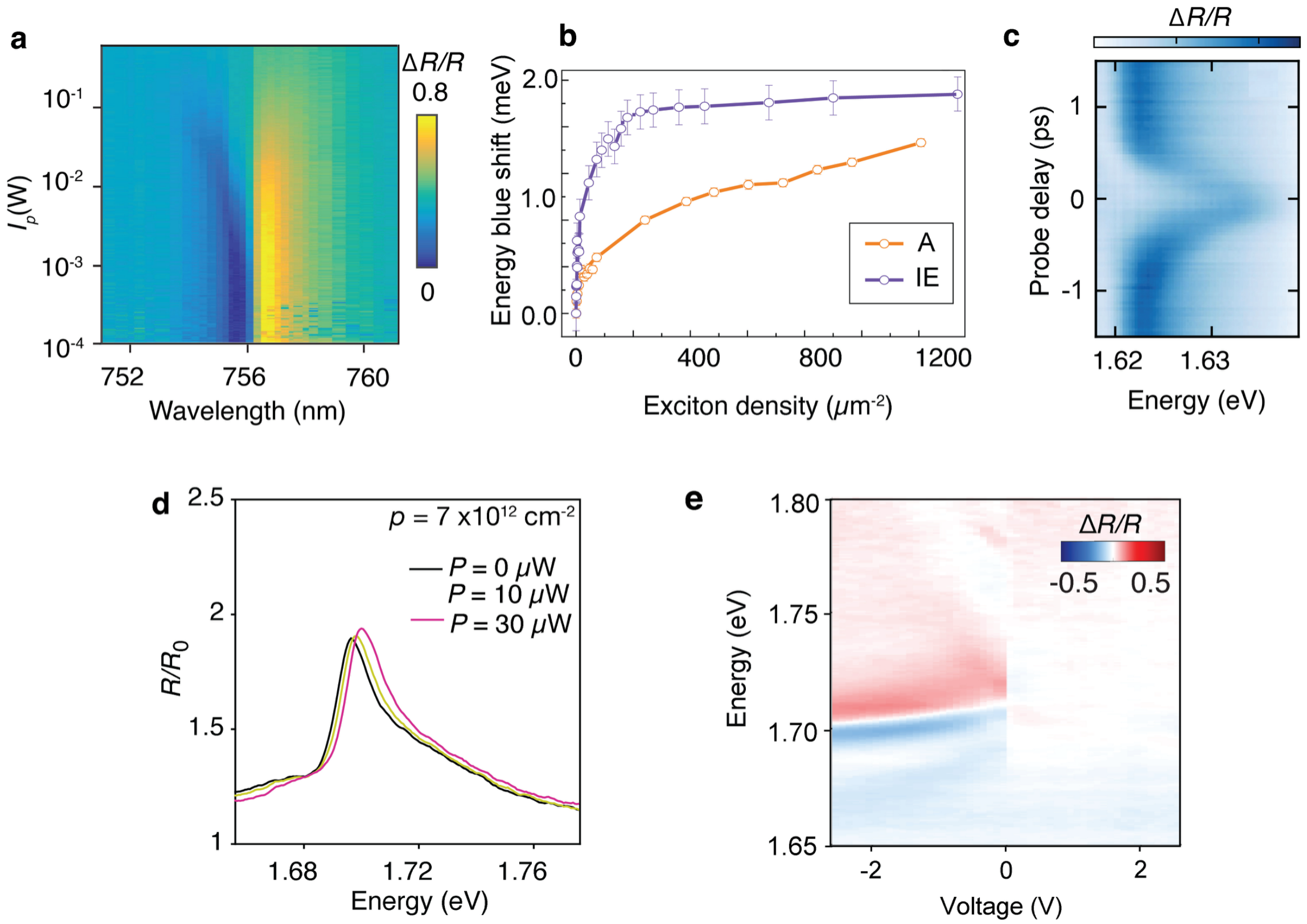

**Figure 8. Exciton and polaron optical nonlinearity in TMD.** **a**, Reflectance of monolayer $MoSe_2$ as a function of the peak laser power at 4K. **b**, Intralayer(A) and interlayer (IE) Exciton reflection relative energy blueshift with increasing exciton density in homobilayer $MoS_2$. **c**, AC stark shift of the attractive polaron in monolayer $MoSe_2$ as a function of the pump-probe delay time under pump laser peak power of 2 GW/cm$^2$. **d**, Trilayer $WSe_2$ Fermi-polaron reflectance contrast as a function of the CW (635nm) laser excitation power. **e**, Relative change in the reflectance induced by CW laser pumping under different doping. The nonlinearity of Fermi polaron only shows up at the hole-doped regime (negative voltage) due to the valley polarization in trilayer $WSe_2$. **a**, Reproduced with permission from *Phys. Rev. Lett*. **120**, 037402 (2018). Copyright 2018 American Physical Society. **b**, Reproduced with permission from *Nat. Commun*. **13**, 6341 (2022). Copyright 2022 Springer Nature Ltd. **c**, Reproduced with permission from *Phys. Rev. Lett*. **132**, 056901 (2024). Copyright 2024 American Physical Society. **d,e**, Reproduced with permission from *Nat. Photon*. **18**, 816–822 (2024). Copyright 2024 Springer Nature Ltd.

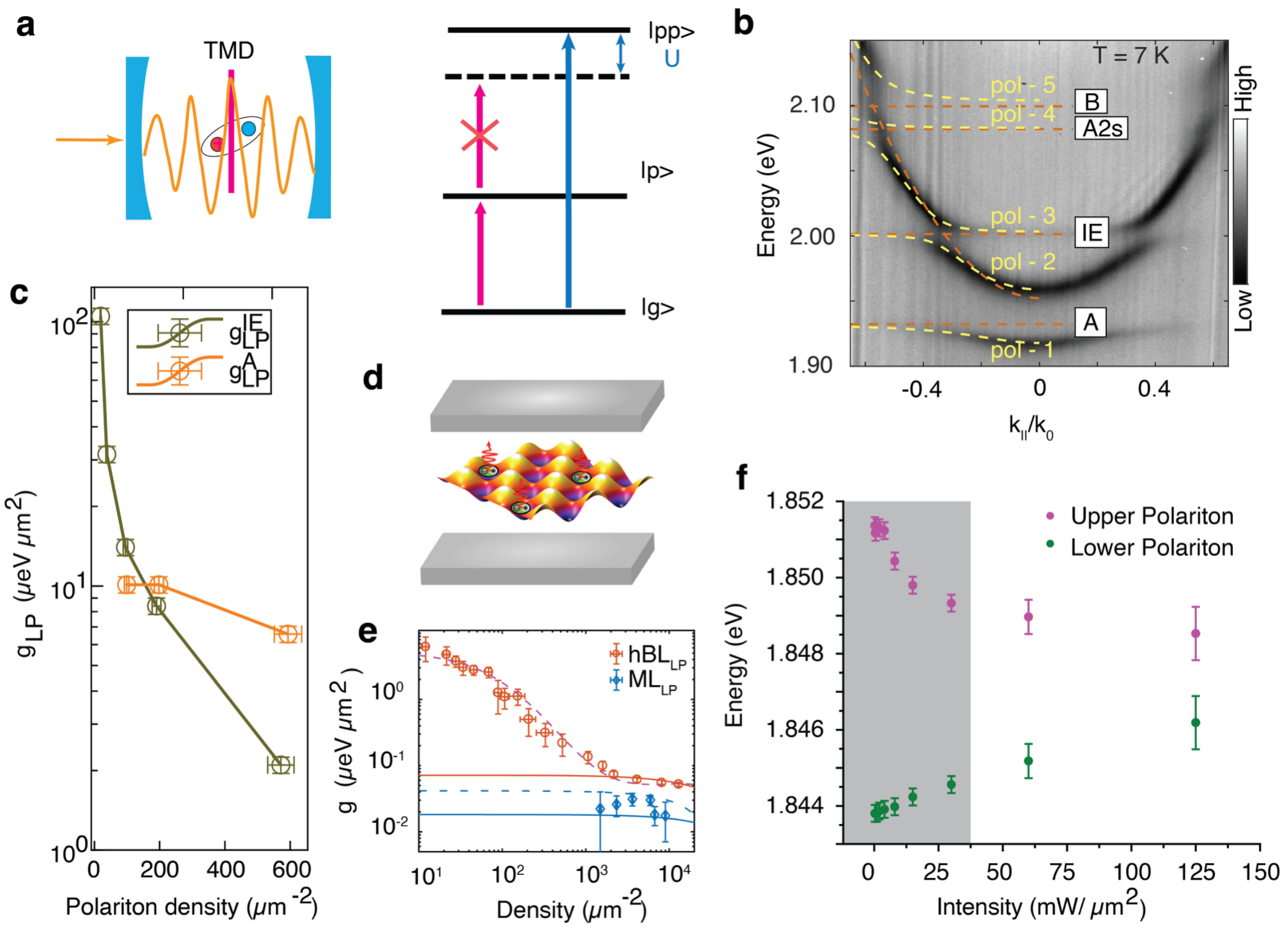

**Figure 9. Nonlinearity enhancement via exciton polariton formation in cavity. a,** Schematic of polariton blockade spectrum happened in cavity. Due to the strong interaction strength U between polaritons (U is larger than the polariton linewidth), two polariton states blueshift. The photon that excites the system from ground to the one-polariton state cannot excite the 2$^{nd}$ polariton anymore due to the addition energy requirement. **b,** Differential reflection at T =7K showing the polariton branches formed by intralayer (A, A 2S, B) and interlayer (IE) exciton coupled to the cavity mode in bilayer $MoS_2$ embedded into microcavity. Strong coupling between the exciton state and cavity mode can be demonstrated by the anti-crossing behavior between each branch. **c**, Nonlinear interaction strength of A and Interlayer exciton as a function of the polariton density. At low polariton density, dipolar polariton owns larger interaction strength due to dipole-dipole repulsion. **d-e**, Moiré polariton system: **d**, Excitons confined in Moiré superlattice and then coupled with cavity photon form moiré polariton, which exhibit large nonlinearity due to the exciton blockade. **e**, Nonlinear interaction strength of moiré lower polariton (orange) and intralayer exciton polariton(blue) as a function of polariton density. **f**, Nonlinearity for 2S polariton. The coupling strength reduces with increasing excitation power density due to the phase space filling. **b, c**, Reproduced with permission from *Nat. Commun*. **13**, 6341 (2022). Copyright 2022 Springer Nature Ltd. **d,e**, Reproduced with permission from *Nature*. **591**, 61–65 (2021). Copyright 2022 Springer Nature Ltd. **f**, Reproduced with permission from *Nat. Commun*. **12**, 2269 (2021). Copyright 2021 Springer Nature Ltd.

Table 1. The second-order nonlinear coefficient for prominent 2d materials

| Material | $\chi^2$ (pm/V) | SHG wavelength(nm) | Thickness | Methods and Substrate |
|---|---|---|---|---|
| $MoSe_2$ | 50 [298] | 810 | 1L | CVD on SiO2 |
| | 7800[297] | 775 | 1L | Exfoliated on Si waveguide |
| | 37[298] | 780 | 1L | Exfoliated on SiO2/Si |
| $MoS_2$ | 5000[134] | 405 | 1L | CVD on SiO2 |
| | 10000[134] | 405 | 1L | Exfoliated on SiO2/Si |
| | 2.2[188] | 780 | Few Layers | Exfoliated on SiO2/Si |
| | 405[76] | 600 | 1L | Exfoliated on SiO2/Si |
| | 5.4[298] | 780 | 1L | Exfoliated on SiO2/Si |
| $WS_2$ | 16.2[298] | 780 | 1L | Exfoliated on SiO2/Si |
| | 4500[299] | 416 | 1L | Exfoliated on SiO2/Si |
| $WSe_2$ | 16.5[298] | 780 | 1L | Exfoliated on SiO2/Si |
| | 5000[300] | 408 | 1L | Exfoliated on SiO2/Si |
| | 100[301] | 775 | 1L | Exfoliated on fused silica |
| h-BN | 42[302] | 406 | 36.2nm | Exfoliated on SiO2/Si |